\documentstyle[12pt]{article}

  \def\pp{{\mathchoice
          {
              \kern 1pt%
              \raise 1pt
              \vbox{\hrule width5pt height0.4pt depth0pt
                    \kern -2pt
                    \hbox{\kern 2.3pt
                          \vrule width0.4pt height6pt depth0pt
                          }
                    \kern -2pt
                    \hrule width5pt height0.4pt depth0pt}%
                    \kern 1pt
           }
            {
              \kern 1pt%
              \raise 1pt
              \vbox{\hrule width4.3pt height0.4pt depth0pt
                    \kern -1.8pt
                    \hbox{\kern 1.95pt
                          \vrule width0.4pt height5.4pt depth0pt
                          }
                    \kern -1.8pt
                    \hrule width4.3pt height0.4pt depth0pt}%
                    \kern 1pt
            }
            {
              \kern 0.5pt%
              \raise 1pt
              \vbox{\hrule width4.0pt height0.3pt depth0pt
                    \kern -1.9pt  
                    \hbox{\kern 1.85pt
                          \vrule width0.3pt height5.7pt depth0pt
                          }
                    \kern -1.9pt
                    \hrule width4.0pt height0.3pt depth0pt}%
                    \kern 0.5pt
            }
            {
              \kern 0.5pt%
              \raise 1pt
              \vbox{\hrule width3.6pt height0.3pt depth0pt
                    \kern -1.5pt
                    \hbox{\kern 1.65pt
                          \vrule width0.3pt height4.5pt depth0pt
                          }
                    \kern -1.5pt
                    \hrule width3.6pt height0.3pt depth0pt}%
                    \kern 0.5pt
            }
        }}

  \def\mm{{\mathchoice
                       {
                             \kern 1pt
               \raise 1pt    \vbox{\hrule width5pt height0.4pt depth0pt
                                  \kern 2pt
                                  \hrule width5pt height0.4pt depth0pt}
                             \kern 1pt}
                       {
                            \kern 1pt
               \raise 1pt \vbox{\hrule width4.3pt height0.4pt depth0pt
                                  \kern 1.8pt
                                  \hrule width4.3pt height0.4pt depth0pt}
                             \kern 1pt}
                       {
                            \kern 0.5pt
               \raise 1pt
                            \vbox{\hrule width4.0pt height0.3pt depth0pt
                                  \kern 1.9pt
                                  \hrule width4.0pt height0.3pt depth0pt}
                            \kern 1pt}
                       {
                           \kern 0.5pt
             \raise 1pt  \vbox{\hrule width3.6pt height0.3pt depth0pt
                                  \kern 1.5pt
                                  \hrule width3.6pt height0.3pt depth0pt}
                           \kern 0.5pt}
                       }}

\catcode`@=11
\def\un#1{\relax\ifmmode\@@underline#1\else
        $\@@underline{\hbox{#1}}$\relax\fi}
\catcode`@=12

\let\du=\du                     

\def\a{\alpha}
\def\b{\beta}

\def\d{\delta}

\def\f{\phi}
\def\g{\gamma}
\def\h{\eta}

\def\j{\psi}
\def\k{\kappa}
\def\l{\lambda}
\def\m{\mu}
\def\n{\nu}

\def\p{\pi}
\def\q{\theta}
\def\r{\rho}
\def\s{\sigma}
\def\t{\tau}

\def\x{\xi}

\def\F{\Phi}
\def\G{\Gamma}

\def\L{\Lambda}
\def\O{\Omega}
\def\P{\Pi}

\def\S{\Sigma}

\def\ve{\varepsilon}

\def\cd{{\cal D}}

\def\cf{{\cal F}}
\def\cg{{\cal G}}

\def\cy{{\cal Y}}
\def\cz{{\cal Z}}

\def\bo{{\raise-.5ex\hbox{\large$\Box$}}}               
\def\pa{\partial}                                       
\def\pr{\prod}                                          
\def\TH{{\raise.2ex\hbox{$\displaystyle \bigodot$}\mskip-4.7mu \llap H \;}}
\def\face{{\raise.2ex\hbox{$\displaystyle \bigodot$}\mskip-2.2mu \llap 
{$\ddot
        \smile$}}}                                      
\def\dg{\sp\dagger}                              

\def\sp#1{{}^{#1}}                              
\def\Tilde#1{\widetilde{#1}}                    
\def\Bar#1{\overline{#1}}                       
\def\VEV#1{\left\langle #1\right\rangle}        
\def\abs#1{\left| #1\right|}                    
\def\leftrightarrowfill{$\mathsurround=0pt \mathord\leftarrow \mkern-6mu
        \cleaders\hbox{$\mkern-2mu \mathord- \mkern-2mu$}\hfill
        \mkern-6mu \mathord\rightarrow$}
\def\dvec#1{\vbox{\ialign{##\crcr
        \leftrightarrowfill\crcr\noalign{\kern-1pt\nointerlineskip}
        $\hfil\displaystyle{#1}\hfil$\crcr}}}           
\def\dt#1{{\buildrel {\hbox{\LARGE .}} \over {#1}}}  

\def\frac#1#2{{\textstyle{#1\over\vphantom2\smash{\raise.20ex
        \hbox{$\scriptstyle{#2}$}}}}}                   
\def\sfrac#1#2{{\vphantom1\smash{\lower.5ex\hbox{\small$#1$}}\over
        \vphantom1\smash{\raise.4ex\hbox{\small$#2$}}}}
\def\bfrac#1#2{{\vphantom1\smash{\lower.5ex\hbox{$#1$}}\over
        \vphantom1\smash{\raise.3ex\hbox{$#2$}}}}       
\def\afrac#1#2{{\vphantom1\smash{\lower.5ex\hbox{$#1$}}\over#2}}    
\def\on#1#2{\mathop{\null#2}\limits^{#1}}               
\def\bvec#1{\on\leftarrow{#1}}                  

\def\[{\lfloor{\hskip 0.35pt}\!\!\!\lceil}
\def\]{\rfloor{\hskip 0.35pt}\!\!\!\rceil}

\def\du#1#2{_{#1}{}^{#2}}
\def\ud#1#2{^{#1}{}_{#2}}

\def\fracm#1#2{\hbox{\large{${\frac{{#1}}{{#2}}}$}}}
\def\ha{{\fracmm12}}
\def\tr{{\rm tr}}

\def\un{\underline}
\def\fracmm#1#2{{{#1}\over{#2}}}

\def\low#1{{\raise -3pt\hbox{${\hskip 0.75pt}\!_{#1}$}}}

\def\Dot#1{\buildrel{_{_{\hskip 0.01in}\bullet}}\over{#1}}
\def\dt#1{\Dot{#1}}

\def\Tilde#1{{\widetilde{#1}}\hskip 0.015in}

\newskip\humongous \humongous=0pt plus 1000pt minus 1000pt
\def\caja{\mathsurround=0pt}
\def\eqalign#1{\,\vcenter{\openup2\jot \caja
        \ialign{\strut \hfil$\displaystyle{##}$&$
        \displaystyle{{}##}$\hfil\crcr#1\crcr}}\,}
\newif\ifdtup

\def\ref#1{$\sp{#1)}$}

\def\pl#1#2#3{Phys.~Lett.~{\bf {#1}B} (19{#2}) #3}
\def\np#1#2#3{Nucl.~Phys.~{\bf B{#1}} (19{#2}) #3}
\def\prl#1#2#3{Phys.~Rev.~Lett.~{\bf #1} (19{#2}) #3}
\def\pr#1#2#3{Phys.~Rev.~{\bf D{#1}} (19{#2}) #3}
\def\cqg#1#2#3{Class.~and Quantum Grav.~{\bf {#1}} (19{#2}) #3}

\def\mpl#1#2#3{Mod.~Phys.~Lett.~{\bf A{#1}} (19{#2}) #3}

\def\ibid#1#2#3{{\it ibid.}~{\bf {#1}} (19{#2}) #3}

\newcommand{\youngA}{
\begin{picture}(30,10)
\linethickness{\unitlength}
\multiput(0,0)(10,0){1}{\framebox(9,9){$\cdot$}}
\end{picture}}

\newcommand{\youngB}{
\begin{picture}(30,10)
\linethickness{\unitlength}
\multiput(0,0)(10,0){2}{\framebox(9,9){$\cdot$}}
\end{picture}}

\newcommand{\youngC}{
\begin{picture}(30,20)
\linethickness{\unitlength}
\multiput(0,10)(10,0){1}{\framebox(9,9){$\cdot$}}
\multiput(0,0)(10,0){1}{\framebox(9,9){$\cdot$}}
\end{picture}}

\newcommand{\youngK}{
\begin{picture}(30,20)
\linethickness{\unitlength}
\multiput(0,10)(10,0){2}{\framebox(9,9){$\cdot$}}
\multiput(0,0)(10,0){1}{\framebox(9,9){$\cdot$}}
\end{picture}}

\newcommand{\youngH}{
\begin{picture}(30,30)
\linethickness{\unitlength}
\multiput(0,20)(10,0){1}{\framebox(9,9){$\cdot$}}
\multiput(0,10)(10,0){1}{\framebox(9,9){$\cdot$}}
\multiput(0,0)(10,0){1}{\framebox(9,9){$\cdot$}}
\end{picture}}

\newcommand{\youngJ}{
\begin{picture}(30,40)
\linethickness{\unitlength}
\multiput(0,20)(10,0){2}{\framebox(9,9){$\cdot$}}
\multiput(0,10)(10,0){1}{\framebox(9,9){$\cdot$}}
\multiput(0,0)(10,0){1}{\framebox(9,9){$\cdot$}}
\end{picture}}

\newcommand{\youngD}{
\begin{picture}(30,40)
\linethickness{\unitlength}
\multiput(0,30)(10,0){1}{\framebox(9,9){$\cdot$}}
\multiput(0,20)(10,0){1}{\framebox(9,9){$\cdot$}}
\multiput(0,10)(10,0){1}{\framebox(9,9){$\cdot$}}
\multiput(0,0)(10,0){1}{\framebox(9,9){$\cdot$}}
\end{picture}}

\newcommand{\youngE}{
\begin{picture}(30,40)
\linethickness{\unitlength}
\multiput(0,30)(10,0){2}{\framebox(9,9){$\cdot$}}
\multiput(0,20)(10,0){2}{\framebox(9,9){$\cdot$}}
\multiput(0,10)(10,0){1}{\framebox(9,9){$\cdot$}}
\multiput(0,0)(10,0){1}{\framebox(9,9){$\cdot$}}
\end{picture}}

\newcommand{\youngI}{
\begin{picture}(30,40)
\linethickness{\unitlength}
\multiput(0,10)(10,0){2}{\framebox(9,9){$\cdot$}}
\multiput(0,0)(10,0){2}{\framebox(9,9){$\cdot$}}
\end{picture}}

\newcommand{\youngR}{
\begin{picture}(30,40)
\linethickness{\unitlength}
\multiput(0,20)(10,0){2}{\framebox(9,9){$\cdot$}}
\multiput(0,10)(10,0){2}{\framebox(9,9){$\cdot$}}
\multiput(0,0)(10,0){2}{\framebox(9,9){$\cdot$}}
\end{picture}}

\newcommand{\youngL}{
\begin{picture}(30,40)
\linethickness{\unitlength}
\multiput(0,20)(10,0){2}{\framebox(9,9){$\cdot$}}
\multiput(0,10)(10,0){2}{\framebox(9,9){$\cdot$}}
\multiput(0,0)(10,0){1}{\framebox(9,9){$\cdot$}}
\end{picture}}

\newcommand{\youngM}{
\begin{picture}(30,40)
\linethickness{\unitlength}
\multiput(0,30)(10,0){2}{\framebox(9,9){$\cdot$}}
\multiput(0,20)(10,0){1}{\framebox(9,9){$\cdot$}}
\multiput(0,10)(10,0){1}{\framebox(9,9){$\cdot$}}
\multiput(0,0)(10,0){1}{\framebox(9,9){$\cdot$}}
\end{picture}}

\newcommand{\youngP}{
\begin{picture}(30,40)
\linethickness{\unitlength}
\multiput(0,30)(10,0){2}{\framebox(9,9){$\cdot$}}
\multiput(0,20)(10,0){2}{\framebox(9,9){$\cdot$}}
\multiput(0,10)(10,0){2}{\framebox(9,9){$\cdot$}}
\multiput(0,0)(10,0){1}{\framebox(9,9){$\cdot$}}
\end{picture}}

\newcommand{\youngF}{
\begin{picture}(30,40)
\linethickness{\unitlength}
\multiput(0,30)(10,0){2}{\framebox(9,9){$\cdot$}}
\multiput(0,20)(10,0){2}{\framebox(9,9){$\cdot$}}
\multiput(0,10)(10,0){2}{\framebox(9,9){$\cdot$}}
\multiput(0,0)(10,0){2}{\framebox(9,9){$\cdot$}}
\end{picture}}

\newcommand{\youngS}{
\begin{picture}(30,10)
\linethickness{\unitlength}
\multiput(0,0)(10,0){1}{\framebox(9,9){}}
\end{picture}}

\newcommand{\youngT}{
\begin{picture}(30,20)
\linethickness{\unitlength}
\multiput(0,10)(10,0){1}{\framebox(9,9){}}
\multiput(0,0)(10,0){1}{\framebox(9,9){}}
\end{picture}}

\newcommand{\youngX}{
\begin{picture}(30,40)
\linethickness{\unitlength}
\multiput(0,30)(10,0){1}{\framebox(9,9){}}
\multiput(0,20)(10,0){1}{\framebox(9,9){}}
\multiput(0,10)(10,0){1}{\framebox(9,9){}}
\multiput(0,0)(10,0){1}{\framebox(9,9){}}
\end{picture}}

\newcommand{\youngY}{
\begin{picture}(30,40)
\linethickness{\unitlength}
\multiput(0,30)(10,0){2}{\framebox(9,9){}}
\multiput(0,20)(10,0){2}{\framebox(9,9){}}
\multiput(0,10)(10,0){1}{\framebox(9,9){}}
\multiput(0,0)(10,0){1}{\framebox(9,9){}}
\end{picture}}

\newcommand{\youngZ}{
\begin{picture}(30,40)
\linethickness{\unitlength}
\multiput(0,30)(10,0){2}{\framebox(9,9){}}
\multiput(0,20)(10,0){2}{\framebox(9,9){}}
\multiput(0,10)(10,0){2}{\framebox(9,9){}}
\multiput(0,0)(10,0){2}{\framebox(9,9){}}
\end{picture}}

\parskip=\medskipamount                 
\thicklines                  

\begin{document}


\thispagestyle{empty}         

\def\border{                                            
        \setlength{\unitlength}{1mm}
        \newcount\xco
        \newcount\yco
        \xco=-24
        \yco=12
        \begin{picture}(140,0)
        \put(-20,11){\tiny Institut f\"ur Theoretische Physik Universit\"at
Hannover~~ Institut f\"ur Theoretische Physik Universit\"at Hannover~~
Institut f\"ur Theoretische Physik Hannover}
        \put(-20,-241.5){\tiny Institut f\"ur Theoretische Physik 
Universit\"at Hannover~~ 
Institut f\"ur Theoretische Physik Universit\"at Hannover~~
Institut f\"ur Theoretische Physik Hannover}
        \end{picture}
        \par\vskip-8mm}

\def\headpic{                                           
        \indent
        \setlength{\unitlength}{.8mm}
        \thinlines
        \par
        \begin{picture}(29,16)
        \put(75,16){\line(1,0){4}}
        \put(80,16){\line(1,0){4}}
        \put(85,16){\line(1,0){4}}
        \put(92,16){\line(1,0){4}}

        \put(85,0){\line(1,0){4}}
        \put(89,8){\line(1,0){3}}
        \put(92,0){\line(1,0){4}}

        \put(85,0){\line(0,1){16}}
        \put(96,0){\line(0,1){16}}
        \put(92,16){\line(1,0){4}}

        \put(85,0){\line(1,0){4}}
        \put(89,8){\line(1,0){3}}
        \put(92,0){\line(1,0){4}}

        \put(85,0){\line(0,1){16}}
        \put(96,0){\line(0,1){16}}
        \put(79,0){\line(0,1){16}}
        \put(80,0){\line(0,1){16}}
        \put(89,0){\line(0,1){16}}
        \put(92,0){\line(0,1){16}}
        \put(79,16){\oval(8,32)[bl]}
        \put(80,16){\oval(8,32)[br]}

        \end{picture}
        \par\vskip-6.5mm
        \thicklines}

\border\headpic {\hbox to\hsize{
\vbox{\noindent AEI -- 097  \hfill December 1998 \\
DESY 98 -- 190 \hfill hep-th/9812051 \\
ITP--UH--32/98 }}

\noindent
\vskip1.3cm
\begin{center}

{\Large\bf Born-Infeld-Goldstone superfield actions
\vglue.1in
for gauge-fixed D-5- and D-3-branes in 6d~\footnote{
Supported in part by the `Deutsche Forschungsgemeinschaft'}}\\
\vglue.3in

Sergei V. Ketov \footnote{
Also at High Current Electronics Institute of Russian Academy of Sciences,
Siberian Branch, \newline ${~~~~~}$ Akademichesky~4, Tomsk 634055, Russia}

{\it Max-Planck-Institut f\"ur Gravitationsphysik}\\
{\it (Albert-Einstein-Institut)}\\ 
{\it Schlaatzweg 1, D--14473 Potsdam, Germany}\\
{\sl ketov@mimir.aei-potsdam.mpg.de}

and

{\it Institut f\"ur Theoretische Physik, Universit\"at Hannover}\\
{\it Appelstra\ss{}e 2, 30167 Hannover, Germany}\\
{\sl ketov@itp.uni-hannover.de}
\end{center}
\vglue.2in
\begin{center}
{\Large\bf Abstract}
\end{center}

The supersymmetric Born-Infeld actions describing gauge-fixed D-5- and
D-3-branes in ambient six-dimensional (6d) spacetime are constructed in 
superspace. A new 6d action is the (1,0) supersymmetric extension of the 
6d Born-Infeld action. It is related via dimensional reduction to another
remarkable 4d action describing the N=2 supersymmetric extension of the 
Born-Infeld-Nambu-Goto action with two real scalars. Both actions are the 
Goldstone actions associated with partial (1/2) spontaneous breaking of 
extended supersymmetry having 16 supercharges down to 8 supercharges. Both 
actions can be put into the `non-linear sigma-model' form by using certain 
non-linear superfield constraints.  The unbroken supersymmetry is always 
linearly realised in our construction. 

\newpage

\section{Introduction}

As an introduction, we remind the reader about some general features of the 
gauge-invariant and gauge-fixed D-brane actions in components, along the lines of
ref.~\cite{aps} (see also refs.~\cite{nil,ber,sor,kal}). This also allows us to specify 
our motivation to introduce superspace in some particular cases. 

The good starting point is provided by a type-II D-p-brane embedded into
flat 10-dimensional (10d) spacetime. The gauge-invariant D-p-brane effective
action is usually written down in terms of the worldvolume fields 
$(X^m(\x),\q_{\a A}(\x),A_{\m}(\x))$ depending upon worldvolume 
coordinates $\x^{\m}$, where $(X^m,\q_{\a A})$ themselves can be 
considered as
the coordinates of N=2 superspace in 10 dimensions $(m=0,1,\ldots,9,~
\a=1,\ldots,16,~ A=1,2)$, whereas $A_{\m}$ is an abelian
gauge field, $\m=0,1,\ldots,p$. The gauge symmetries of the action comprise 
(i) worldvolume diffeomorphisms, (ii) a fermionic $\k$-symmetry, and (iii)
a $U(1)$ gauge invariance, whereas the global or rigid invariances are given 
by 10d, N=2 super-Poincar\'e symmetry. The gauge-invariant D-p-brane action 
is a sum of the Born-Infeld-Nambu-Goto (BING) and Wess-Zumino (WZ) 
terms,~\footnote{The D-brane torsion coefficient in front of the action 
is chosen to be one.}
$$ S_p = - \int d^{p+1}\x \sqrt{-\det(\cg_{\m\n}+\cf_{\m\n})}
+\int \O_{p+1}~,\eqno(1.1)$$
where $\cg_{\m\n}$ is the supersymmetric induced metric in the 
worldvolume,
$$ \cg_{\m\n}=\h_{mn}\P^m_{\m}\P^n_{\n}~,\quad
\P^m_{\m}=\pa_{\m}X^m-\bar{\q}\G^m\pa_{\m}\q~,\eqno(1.2)$$
$\cf_{\m\n}$ is the supersymmetric abelian field strength,
$$ \cf_{\m\n}=\left[ \pa_{\m}A_{\n}-\bar{\q}\hat{\G}\G_m\pa_{\m}\q
\left(\pa_{\n}X^m-\frac{1}{2}\bar{\q}\G^m\pa_{\n}\q\right)\right]
-(\m\leftrightarrow\n)~,\eqno(1.3)$$
and
$$ \hat{\G}=\left\{ \begin{array}{cc} I \otimes\t_3, & 
 ~~p~~ {\rm odd}~, \\
\G_{11}\otimes I, & ~~p~~ {\rm even}~,\end{array}\right.
\eqno(1.4)$$ 
with respect to the $(\a A)$ indices. The WZ term in eq.~(1.1)
describes a coupling of the D-brane to the background Ramond-Ramond
(RR) gauge fields \cite{polbook}, while its explicit form is fixed by
the $\k$-symmetry of the whole action (1.1),
$$ \d_{\k}X^m=\bar{\q}\G^m\d\q~,\quad  \d_{\k}\q=\frac{1}{2}(1+\G)\k~,
\eqno(1.5)$$
where $\G$ is a (field-dependent) projector \cite{aps}. The worldvolume
diffeomorphisms (i) ensure that only the $(9-p)$ coordinates 
$\{X^{i}\}$, $i=p+1,\ldots,9$, transverse to the D-brane worldvolume are 
physical, whereas the $\k$-symmetry (ii) effectively eliminates half of the 
fermionic $\q$'s in accordance with the
BPS nature of the D-brane that  breaks just half of spacetime
supersymmetry. The rigid 10d, N=2 supersymmetry transformations are
$$ \d_{\ve}X^m=\bar{\ve}\G^m\q~,\quad \d_{\ve}\q=\ve~.\eqno(1.6)$$

All physical fields in the D-brane worldvolume can be interpreted as the
{\it Goldstone} fields associated with the symmetries broken by the D-brane 
\cite{polbook,town}. These spontaneously broken symmetries (including 
broken supersymmetry) are therefore to be non-linearly realised in the 
gauge-fixed D-brane action to be obtained by fixing the local symmetries 
and removing unphysical degrees of freedom. A covariant physical gauge for 
the worldvolume general coordinate transformations is given by the so-called
{\it  static} gauge, in which the first $(p+1)$ spacetime 
coordinates are identified with the D-brane worldvolume coordinates, i.e. 
$X^{\m}=\x^{\m}$. The remaining scalars $X^i$ representing transverse 
excitations of the D-brane can then be identified with the Goldstone bosons
(collective modes) $\f^i$ associated with spontaneously broken 
translations \cite{polbook,town}. 
The bosonic part of the induced metric in the static gauge reads
$$ G_{\m\n} = \h_{\m\n} + \pa_{\m}\f^i\pa_{\n}\f^i~,\quad {\rm where}\quad
i=1,\ldots,9-p.~ \eqno(1.7)$$
A covariant gauge-fixing of the $\k$-symmetry is also possible (e.g. taking
either $\q_{\a 1}=0$ and $\q_{\a 2}=\j$ in the type-IIB case, or just the 
opposite, $\q_{\a 1}=\j$ and $\q_{\a 2}=0$, in the type-IIA case), while the
WZ term vanishes in this gauge \cite{aps}. The covariant gauge-fixed D-p-brane
action can therefore be identified with a supersymmetric extension of the 
following BING (or Goldstone)-type action:
$$ S_{\rm bosonic} = - \int d^{p+1} \x \sqrt{-\det\left( \h_{\m\n}+F_{\m\n}+
\pa_{\m}\f^i\pa_{\n}\f^i\right) }~.\eqno(1.8)$$
This action depends upon the abelian gauge field $A_{\m}$ only via its 
field  strength $F_{\m\n}$ so that the $U(1)$ gauge invariance is kept. 

The number $(p-1)+(9-p)=8$ of the bosonic physical degrees of freedom in the
action (1.8) matches with the number of fermionic degrees of freedom
$16/2=8$ associated with the 10d Majorana-Weyl (MW) spinor $\j$, and it 
does not depend upon $p$. It is not, therefore, surprising that 
supersymmetric 
extensions of all the gauge-fixed D-p-brane actions (1.8) can be deduced by 
dimensional reduction from a single master 10d action \cite{aps},
$$ S_{\rm master}= - \int d^{10}\x \sqrt{-\det\left[\h_{\m\n}+F_{\m\n}
-2\bar{\j}\G_{\m}\pa_{\n}\j +(\bar{\j}\G^{\r}\pa_{\m}\j)
(\bar{\j}\G_{\r}\pa_{\n}\j)\right]}~,\eqno(1.9)$$
associated with the top value $p=9$ of the 10d {\it `spacetime-filling'} D-9-brane. 

By construction \cite{aps}, the component 10d super-Born-Infeld (sBI) action 
(1.9) 
is invariant under two 10d MW supersymmetries, one unbroken and another one 
spontaneously broken, with the 10d Maxwell supermultiplet $(A_{\m},\j_{\a})$ 
being the Goldstone vector supermultiplet associated with the second 
non-linearly realised supersymmetry.~\footnote{The 10d supersymmetry 
transformation laws are given in ref.~\cite{aps}.} In particular, the spinor
superpartner $\j_{\a}$ of the BI vector is the Goldstone fermion~\cite{av}.
It should, however, be emphasized that the first unbroken supersymmetry of
the action (1.9) is not the same as the original rigid supersymmetry (1.6)
since it has to be supplemented by the compensating gauge transformation 
needed to preserve the gauge. In other words, neither of supersymmetries is
manifest in the action (1.9). 

Our goal in this paper is to rewrite some of the supersymmetric gauge-fixed 
D-p-brane actions in superspace, in order to make their unbroken 
supersymmetries  manifest. The superfield formulation is useful in deciphering
the unique non-trivial geometry underlying the complicated Goldstone actions 
associated with partial supersymmetry breaking in various spacetime 
dimensions (see e.g., refs.~\cite{bagger,bik} for a recent account of 
non-linear realizations of supersymmetry). The superspace formulation becomes 
indispensable if one wants to address quantum 
properties of D-branes, e.g. their black-hole applications \cite{kle}. 

Supersymmetrizing the BI actions in various spacetime dimensions represents 
a challenge in supersymmetry since one has to deal with a non-polynomial 
field theory containing higher derivatives of all orders. Causal propagation
of the physical fields is to be maintained, while the auxiliary fields 
needed to close the off-shell supersymmetry algebra are to be kept 
non-propagating (the last consistency condition was called the 
`auxiliary freedom' in ref.~\cite{agat}). Both requirements are non-trivial in 
supersymmetric field theories with higher derivatives. As was demonstrated
e.g., in ref.~\cite{swer}, the naive approach based on insisting on 
purely {\it algebraic} equations of motion for the auxiliary fields rules out 
a supersymmetrization of the 4d BI action at all. In fact, it is possible to 
avoid propagating auxiliary fields (i.e to achieve the auxiliary freedom) 
by imposing less restrictive conditions in some particular cases, with the sBI 
actions being the most important examples. It turns out to be possible due to 
the very special (Goldstone) nature of the sBI superfield actions whose physical 
bosonic part is free of ghosts and, hence (if consistent), the auxiliary fields 
should be non-propagating.

Yet another important asset of the BI action in 4d \cite{bi} is its {\it
electric-magnetic} (e.-m.) self-duality \cite{schr} (see also ref.~\cite{gzu}).
 The self-duality and causal propagation \cite{pleb} together are responsible 
for the characteristic (`square root of a determinant') non-polynomial 
structure of the 4d BI action \cite{gr}. It is worth mentioning here that the 
fundamental motivation in favor of the non-linear BI generalization of the 
Maxwell electrodymanics is the well-known BI taming of Coulomb self-energy, i.e.
the existence of a non-singular charged soliton with finite self-energy
\cite{bi,pleb}. 

The leading term in the expansion of the BI action with respect to the gauge
field strength is the Maxwell action. As is well-known, even a covariant 
off-shell manifestly N-extended supersymmetrization of the 4d free (!) 
Maxwell theory is the difficult problem once a number (N) of supersymmetries 
exceeds two. The infinite number of auxiliary fields beyond the 
N=2 (or 8 supercharges) barrier is, in fact, required. In this paper we 
restrict ourselves to the cases of N=1 and N=2 supersymmetry in 4d, and, most 
notably, (1,0) supersymmetry in 6d too, where an off-shell formulation of the 
super-BI theory is still possible in the conventional superspace with finite 
number of auxiliary fields. Similar reasoning (for example, the need for an
off-shell extended superspace formulation of a Fayet-Sohnius hypermultiplet)
also restricts the number of real Goldstone bosons $\f^i$ in the 4d  
(gauge-fixed) super-BING action (1.8), if one wants to achieve its off-shell 
superspace reformulaiton by using a finite set of auxiliary fields. In 4d (i.e 
for a D-3-brane) and N=2 unbroken supersymmetry we are thus led to restrict 
$i=1,2$, which implies the six-dimensional ambient spacetime for the D-3-brane 
to propagate.~\footnote{This supermembrane was first considered in ref.~\cite{hlp}.} 
Accordingly, in 6d we are going to restrict ourselves to the 6d `spacetime-filling' 
D-5-brane whose bosonic gauge-fixed action is the 6d Born-Infeld action. 

It is worthy to be mentioned that the initial motivation to supersymmetrize
the BI action came from the fact that it is the relevant part of the 10d 
open superstring effective action \cite{strings}. In addition, the quartic 
terms in the expansion of the 4d BI action amount to the so-called 
{\it Euler-Heisenberg} (EH) action \cite{eh}, which is known to be the one-loop 
bosonic contribution to the low-energy effective action of supersymmetric 
scalar QED.  
 
Taken together, the above reasoning provides broad and compelling 
motivation for a construction of supersymmetric BI and BING actions in 
superspace, in terms of constrained extended superfields capable to unify 
Goldstone scalars and vectors. The importance of those problems, their 
actuality, as well as some outstanding technical difficulties, related to highly
non-trivial extensions of the known N=1 supersymmetric Maxwell-Goldstone action
\cite{bg} to the sBI actions with extended unbroken supersymmetry, were 
recently emphasized from various points of view in refs.~\cite{bagger,bik,rte,stony}.

We adopt the most straightforward (bottom-up) approach to supersymmetrize the bosonic BI 
action by employing extended superspace, without using the standard coset 
construction underlying non-linear realizations of internal and spacetime
symmetries, including supersymmetry \cite{cwz}. Though  being quite powerful, 
the general theory of non-linear realizations usually leads in practice to
highly involved perturbative calulations in order to arrive at a closed form of
the Goldstone action associated with partially broken supersymmetry. Moreover, 
the coset construction of non-linearly realised supersymmetry turns out to be 
incomplete since it does not automatically imply the irreducibility constraints
on the Goldstone superfields \cite{bagger,bik}. Just using the basic fact that,
being of the Goldstone origin, the bosonic BI action is the unambiguous
consequence of non-linearly realised (broken) supersymmetry, its minimal and
unique completion with respect to unbroken supersymmetry can be most easily 
obtained in appropriate superspace by taking a massless vector supermultiplet 
as the Goldstone one.

The paper is organized as follows: in sect.~2 we review the 4d bosonic 
BI action and then discuss its supersymmetric generalizations, namely, 
(i) the 4d Goldstone action associated with N=2 supersymmetry spontaneously 
broken to N=1 and a massless N=1 vector superfield as the Goldstone-Maxwell
superfield \cite{bg,rte}, and (ii) the 4d Goldstone action associated with N=4 
supersymmetry spontaneously broken to N=2 with a massless N=2 vector superfield
as the Goldstone-Maxwell superfield \cite{k2}. Both supersymmetric BI actions
can be equally interpreted as the gauge-fixed actions of a D-3-brane either
`filling' 4d spacetime or propagating in six-dimensional spacetime, 
respectively. Our main new construction that generalizes those of sect.~2 is
presented in sect.~3, where we formulate for the first time the manifestly 6d 
Lorentz invariant and (1,0) supersymmetric Goldstone action associated with 
partial breaking of (2,0) supersymmetry down to (1,0) supersymmetry in 6d,
with a massless (1,0) vector superfield being the Goldstone-Maxwell superfield 
in 6d. The new action is simultaneously the (1,0) supersymmetric gauge-fixed 
6d `spacetime-filling' D-5-brane action in 6d superspace. Our conclusions are 
summarized in sect.~4.
\vglue.2in

\section{4d (super)BI actions in N=0,1 and 2 superspace} 

In this section we only discuss {\it four}-dimensional supersymmetric BI 
actions, both in components and in superspace. We briefly review some features
of the bosonic BI action, which are going to be relevant for us in what follows.
Then we introduce the N=1 supersymmetric BI action \cite{bg} and generalize it 
further to the N=2 BING action in 4d, N=2 superspace.

\subsection{The bosonic BI action}

The BI action in flat four-dimensional (4d) spacetime with Minkowski metric
$\h_{\m\n}={\rm diag}(+,-,-,-)$,~\footnote{Our notation in this paper differ 
from that of ref.~\cite{k2}.}
$$ S_{\rm BI} = -\fracmm{1}{b^2}\int d^4x\,\sqrt{-\det(\h_{\m\n}+bF_{\m\n})}~,
\eqno(2.1)$$
was introduced \cite{bi} as the non-linear generalization of Maxwell
electrodynamics. This action also naturally arises (i) as the bosonic part of
the 4d low-energy effective action of open superstrings (together with 
other massless superstring modes), and (ii) as the bosonic 4d 
spacetime-filling D-3-brane action as well (sect.~1). In string/brane theory 
$b=2\p\a'$, whereas we choose $b=1$ for notational simplicity.

The BI action (2.1) is manifestly Lorentz-invariant, it depends upon the 
gauge field $A_{\m}$ only via its field strength $F_{\m\n}=\pa_{\m}A_{\n}-
\pa_{\n}A_{\m}$, it contains no spacetime derivatives of $F$, and, after 
being expanded in powers of $F$, it gives the
Maxwell action as the leading contribution. In fact, the BI action shares
with the Maxwell action some other physically important properties, such as
\begin{itemize}
\item causal propagation (no ghosts),
\item positive energy density,
\item electric-magnetic self-duality,
\end{itemize}
which are non-trivial in the BI case \cite{pleb,gr}. Unlike the Maxwell action,
the BI action provides a natural taming of the Coulomb self-energy, which is
yet another argument in favor of quantum consistency of superstring theory!

Taking advantage of the Lorentz invariance of the BI action, it is always 
possible to simplify a calculation of its expansion in powers of the gauge
field strength by putting $F_{\m\n}$ into a particular form, e.g.
$$F_{\m\n}=\left(\begin{array}{cccc} 0 & \l_1 & 0 & 0 \cr
-\l_1 & 0 & 0 & 0 \cr 0 & 0 & 0 & \l_2 \cr 0 & 0 & -\l_2 & 0 \cr
\end{array}\right)\eqno(2.2)$$
in terms of real `eigenvalues' $(\l_1,\l_2)$. Eq.~(2.2) is, of course, just a
manifestation of the fact that the Lorentz group $SO(1,3)$ has merely two 
independent Casimir operators. In other words, it suffices to pick up two
independent Lorentz-invariant $F$-products in order to parametrize any 
Lorentz-invariant function of $F_{\m\n}$. For example,
$$  \det(\h_{\m\n}+ F_{\m\n}) = -1 - \frac{1}{2}F^2 +\det(F_{\m\n})=
 -1 - \frac{1}{2}F^2  +\frac{1}{4}\left[F^4- \frac{1}{2}(F^2)^2\right]~,
\eqno(2.3)$$
where we have introduced two real independent Lorentz-invariants as follows:
$$ F^2\equiv F^{\m\n}F_{\m\n}\quad {\rm and}\quad
F^4\equiv F_{\m\n}F^{\n\l}F_{\l\r}F^{\r\m}~.\eqno(2.4)$$
The choice (2.4) is, of course, not unique, and it is not really the most
convenient one in 4d supersymmetry. So let's introduce the 4d dual of 
$F_{\m\n}$,
$$ \tilde{F}^{\m\n}= \frac{1}{2}\ve^{\m\n\l\r}F_{\l\r}~,\eqno(2.5)$$
and form (anti)self-dual linear combinations,
$$ F^{\pm}\low{\m\n} = \frac{1}{2}\left( F\pm i\tilde{F}\right)_{\m\n}~,
\eqno(2.6)$$
that satisfy the identities  
$$ (F^{\pm})^2 = \frac{1}{2}(F^2\pm iF\tilde{F})~,\quad 
(F^2)^2+(F\tilde{F})^2=4(F^+)^2(F^-)^2~.\eqno(2.7)$$
Note that
$$ \det(F_{\m\n})= \frac{1}{16}(F\tilde{F})^2~.\eqno(2.8)$$

Using yet another identity
$$F^4= \frac{1}{2}(F^2)^2 + \frac{1}{4}(F\tilde{F})^2~,\eqno(2.9)$$
it is possible to slightly simplify eq.~(2.3) to the form
$$  -\det(\h_{\m\n}+ F_{\m\n}) = 1 + \frac{1}{2}F^2 
 - \frac{1}{16}(F\tilde{F})^2~,\eqno(2.10)$$
which implies
$$ \eqalign{
L_{\rm BI} \equiv 1-\sqrt{ -\det(\h_{\m\n}+ F_{\m\n})}
&  = 
-\frac{1}{4}F^2 -\frac{1}{8}\left[ \frac{1}{4}(F^2)^2-F^4\right]+O(F^6) \cr 
& = -\frac{1}{4}F^2 +
\frac{1}{32}\left[ (F^2)^2+(F\tilde{F})^2\right] + O(F^6) \cr
& =  -\frac{1}{4}F^2  +\frac{1}{8}(F^+)^2(F^-)^2+O(F^6)~.\cr}
\eqno(2.11)$$

By a complex `rotation' of Lie algebra of $SO(1,3)$ to that of 
$SL(2,{\bf C})$, it is sometimes useful (in supersymmetry) to replace 
$F^+_{\m\n}$ by a $2\times 2$ matrix
$$ \hat{F}\du{\a}{\b}=(\s^{\m\n})\du{\a}{\b}F_{\m\n}~,\quad \a,\b=1,2~,
\eqno(2.12)$$
where we have introduced the two-component spinor notation,
$$ (\s^{\m\n})=\frac{1}{4}\left(\s^{\m}\tilde{\s}^{\n}
-\s^{\n}\tilde{\s}^{\m}\right),\quad \s^{\m}=({\bf 1},\vec{\s}), \quad
\tilde{\s}^{\m}=({\bf 1},-\vec{\s})~,\eqno(2.13)$$
in terms of Pauli matrices $\vec{\s}$. We find in addition that
$$ \frac{1}{4}\abs{\det\hat{F}}^2=4(F^+)^2(F^-)^2=(F^2)^2+(F\tilde{F})^2~,
\eqno(2.14)$$
where we have introduced the chiral $(2\times 2)$ determinant, $\det\hat{F}$, on the 
left-hand-side. The right-hand-side of the identity (2.14) is often referred to as the 
{\it Euler-Heisenberg} (EH) lagrangian \cite{eh}. It arises, 
in particular, as the bosonic part of the one-loop effective action in N=1 
supersymmetric scalar electrodynamics (= the supersymmetric quantum field 
theory of a massive N=1 scalar multiplet minimally coupled to an N=1 vector 
multiplet in 4d) with the parameter  $b^2=e^4/(24\p^2m^4)$. 

The single complex Lorentz invariant 
$$ \frac{1}{16}\tr(\hat{F}^2)=-\frac{1}{4}F^2-\frac{i}{4}F\tilde{F}\equiv
A + i B \eqno(2.15)$$
is another natural variable for an expansion of the BI action in terms of
the field strength $F$ (it will be used in subsect.~2.2). Yet another choice 
of variables to be used in subsect.~2.3 is given by the Maxwell lagrangian and
the Maxwell energy-momentum tensor squared (= the EH lagrangian!),
$$ -\frac{1}{4}F^2=A \quad {\rm and} \quad \frac{1}{32}\left[ (F^2)^2+
(F\tilde{F})^2\right]\equiv E~.\eqno(2.16)$$
The Lorentz invariants (2.16) have natural supersymmetric extensions 
(subsect.~2.2 and 2.3), with the first one having the form of a {\it chiral} 
superspace integral while the second one being a {\it full} superspace 
integral. This justifies our choice (2.16). We find
$$ -\det(\h_{\m\n}+F_{\m\n})=1-2A-B^2=(1-A)^2 -2E~.\eqno(2.17)$$
This allows us to rewrite the BI lagrangian to the form
$$ L_{\rm BI}(F)=A+E+\ldots = A+EY(A,E)~,\eqno(2.18)$$
where the function $Y(A,E)$ has been introduced. It is not difficult to check
that $Y(A,E)$ is just a solution to the quadratic equation
$$ Ey^2 + 2(A-1)y+2=0~.\eqno(2.19)$$
Similarly, it is straightforward to calculate $L_{\rm BI}(F)$ as a function
of $A$ and $B$ (see  subsect.~2.2), e.g., by using the identity $A^2+B^2=2E$,
and eqs.~(2.18) and (2.19).

A lagrangian `magnetically dual' to the BI one is obtained via a first-order
action 
$$ L_1=L_{\rm BI}(F) + 
\frac{1}{2}\Tilde{A}_{\m}\ve^{\m\n\l\r}\pa_{\n}F_{\l\r}~,\eqno(2.20)$$
where $\Tilde{A}_{\m}$ is a (dual) magnetic vector potential. 
$\Tilde{A}_{\m}$ enters eq.~(2.20) as the Lagrange multiplier enforcing the 
Bianchi identity $\ve^{\m\n\l\r}\pa_{\n}F_{\l\r}=0$. Varying eq.~(2.20) with 
respect to $F_{\m\n}$ instead, solving the arising algebraic equation on 
$F_{\m\n}$ as a function of the magnetically dual gauge field strength 
${}^*F_{\m\n}=\pa_{\m}\Tilde{A}_{\n}-\pa_{\n}\Tilde{A}_{\m}$ (use the 
representation (2.2) for $F_{\m\n}$ and similarly for ${}^*F_{\m\n}$~!), 
and substituting a solution back into eq.~(2.20) yields the magnetically dual
action in terms of ${}^*F_{\m\n}$, which has {\it the same} form as the 
original BI action (2.1) in terms $F_{\m\n}$. This is called  
{\it electric-magnetic} (e.-m.) self-duality \cite{schr,gzu,gr}, and it is 
connected to the classical $SL(2,{\bf R})$ symmetry of IIB superstrings 
\cite{tse2}. The non-gaussian BI lagrangian (2.1) is {\it uniquely} fixed 
by the requirements of causal propagation and classical e.-m. self-duality 
if, in addition, one insists on the Maxwell low-energy limit, i.e. the (strong) 
correspondence principle. In general, there exists a family of e.-m. 
self-dual lagrangians parametrized by one variable, with all of them being 
solutions to a first-order Hamilton-Jacobi partial differential equation 
\cite{gr}. 

The classical BI action can be rewritten to many equivalent forms by introducing some
auxiliary fields that allow one to get rid of the square root or the determinant in the
action. For instance, it is possible to put the BI action to a classically equivalent form 
that is quadratic in the gauge field strength \cite{hz,rte}. We are not
going to use this kind of tricks in what follows.

\subsection{N=1 sBI action}

The manifestly N=1 supersymmetric 4d Born-Infeld (or Goldstone-Maxwell) 
action associated with partial spontaneous breaking of rigid N=2 supersymmetry in terms of the 
Goldstone-Maxwell N=1 supermultiplet $(A_{\m},\j_{\a},D)$ was constructed in superspace in 
refs.~\cite{cf,bg} (see also ref.~\cite{rte}). Amongst the superpartners of the Maxwell gauge 
field $A_{\m}$ are the Goldstone (Majorana) fermion $\j_{\a}$ and the real 
auxiliary scalar $D$. In this subsection we briefly review some of the 
results of ref.~\cite{bg}, since the N=1 supersymmetric Goldstone-Maxwell action 
provides the basic pattern that will be subsequently generalized to extended unbroken supersymmetry 
in the next subsect.~2.3.
  
The standard 4d, N=1 superspace is parametrized by the coordinates 
$Z^M=(x^{\m},\q\low{\a},\bar{\q}_{\dt{\a}})$, where $\q\low{\a}$ and 
$\bar{\q}_{\dt{\a}}$ are (Majorana) spinor anticommuting coordinates in the 
2-component notation, $(\q\low{\a})^*=\bar{\q}_{\dt{\a}}$ and $\a=1,2$. An
abelian vector N=1 supermultiplet is described in N=1 superspace by the 
irreducible chiral spinor superfield $W_{\a}$ satisfying the off-shell 
constraints \cite{bw}
$$ \bar{D}_{\dt{\a}}W\low{\a}=0~,\quad D^{\a}W\low{\a}=\bar{D}_{\dt{\a}}
\bar{W}^{\dt{\a}}~.\eqno(2.21)$$
As a result of these constraints, the bosonic components of the N=1 superfield
strength $W_{\a}$ can be introduced as follows \cite{bw}:
$$ \left. D^{\a}W_{\b}\right| = (\s^{\m\n})\du{\b}{\a}F_{\m\n}
+i\d^{\a}_{\b}D~,\eqno(2.22)$$
where $F_{\m\n}$ is the Maxwell field strength of the gauge field $A_{\m}$. 

The superfield constraints (2.21) can be solved in terms of a real gauge
superfield pre-potential $V(x,\q,\bar{\q})$ as \cite{bw}
$$ W_{\a}=\bar{D}^2D_{\a}V~,\eqno(2.23)$$
subject to abelian gauge transformations $\d V=i(\L-\bar{\L})$ where $\L$ is
a chiral superfield gauge parameter, $\bar{D}_{\dt{\a}}\L=0$. This gives the 
necessary input for a superfield quantization in terms of the unconstrained 
superfield $V$. 

The 4d, N=2 supersymmetry algebra can be decomposed with respect to unbroken
N=1 supersymmetry as \cite{bg}
$$\eqalign{
\{Q\low{\a},\bar{Q}_{\dt{\a}}\}=2\s^{\m}_{\a\dt{\a}}P_{\m}~,\quad & \quad 
 \{S\low{\a},\bar{S}_{\dt{\a}}\}=2\s^{\m}_{\a\dt{\a}}P_{\m}~,\cr
\{Q\low{\a},S\low{\b}\}=0~,\quad & \quad \{Q\low{\a},\bar{S}_{\dt{\a}}\}=0~,
\cr}\eqno(2.24)$$
where $Q$'s stand for the unbroken (N=1) supersymmetry generators, $S$'s stand for the broken 
(N=1) supersymmetry generators, while $P_{\m}$ are 4d translation generators. It is worth
mentioning that a non-vanishing central charge does not appear in the N=2 algebra (2.24). 
The vanishing central charge is, in fact, required for a consistency of the 
Goldstone-Maxwell action. In particular, the BPS nature of this action also implies the
vanishing vacuum expectation values for composites of the physical fields, while the auxiliary field 
$D$ should vanish on-shell too. In general, vanishing vacuum expectations for the physical (Goldstone) 
composites protect the auxiliary fields from becoming propagating due to interacting terms in all 
suppersymmetric BI actions provided that pure kinetic terms for the auxiliary fields do not appear. The
latter turns out to be the case for the manifestly supersymmetric (superfield) Born-Infeld-Goldstone 
actions considered in this paper. 

A generalization of the N=1 supersymmetric constraints (2.21), which would be 
invariant under the second $(S)$ non-linearly realised supersymmetry, is
possible, in principle, by using the standard perturbative approach of
non-linear realizations \cite{cwz}, though the full answer in a closed form
is still unknown in this case~\cite{bg}. It is, nevertheless, possible to
determine the full and manifestly N=1 supersymmetric Goldstone-Maxwell action
by a direct (and unique) N=1 supersymmetrization of the BI action, as in
ref.~\cite{cf}. The result is given by the sBI action \cite{cf,bg,rte}
$$ \eqalign{
S\low{\rm N=1~GM} &=~ \left[ \frac{1}{4}\int d^4x d^2\q \, W^2 +{\rm h.c.} 
\right]  + \frac{1}{8} \int d^4xd^2\q d^2\bar{\q} \, f(A,B)W^2\bar{W}^2 \cr
&=~  \frac{1}{4}\int d^4x d^2\q \left\{ W^2+\frac{1}{4}\bar{D}^2\left[
f(A,B)W^2\bar{W}^2\right]\right\} +{\rm h.c.} \cr
&\equiv~  \frac{1}{4}\int d^4x d^2\q \, W^2_{\rm improved}  +{\rm h.c.}~, 
\cr} \eqno(2.25)$$
where the structure function $f(A,B)$ is given by
$$ f(A,B)= \fracmm{1}{1-A+\sqrt{1-2A -B^2}}~~,\eqno(2.26)$$
whereas $A$ and $B$ stand for the N=1 superfields
$$ \eqalign{
A =  \frac{1}{4}D^2W^2 + {\rm h.c.}~,\cr
iB  = \frac{1}{4}D^2W^2 - {\rm h.c.}~,\cr}\eqno(2.27)$$
respectively, whose leading ($F$-dependent) components 
(at $\q\low{\a}=\bar{\q}_{\dt{\a}}=0$) are just given by $A$ and $B$ of
eq.~(2.15).~\footnote{It is customary (in supersymmetry) to denote both a
superfield and its first component by the \newline ${~~~~~}$ same letter. 
This slight abuse of notation, hopefully, does not lead to a confusion.} 

The Goldstone-Maxwell action (2.25) is thus given by a sum of 
the chiral N=1 superspace integral (= super-Maxwell or super-$A$ invariant) 
and the full N=1 superspace integral (= super Euler-Heisenberg or super-$E$
invariant), with the latter being modified by the `formfactor' 
$f(D^2W^2,\bar{D}^2\bar{W}^2)$. The only quartic (higher derivative) 
combination, $\frac{1}{4}(F^2)^2-F^4$, that can be supersymmetrized up to the 
full (EH) N=1 superinvariant, was earlier identified in ref.~\cite{di} by using 
helicity conservation of four-particle scattering amplitudes in N=1
supersymmetric scalar QED.

In terms of our `smart' variables (2.16) the bosonic BI lagrangian in the form
(2.18) can be immediately supersymmetrized to the form (2.25). The N=1 supersymmetric
Goldstone-Maxwell action is therefore given by the N=1 supersymmetric Born-Infeld action. As was
recently argued in ref.~\cite{rte}, the {\it same} sBI action emerges from the N=2 supersymmetric 
non-linear {\it APT model} \cite{apt}, where N=2 supersymmetry is partially broken to N=1 supersymmetry
due to the non-linearity of the Seiberg-Witten-type action for an N=2 vector supermultiplet in the 
presence of `electric' {\it and} `magnetic' {\it Fayet-Iliopoulos} (FI) terms, after `integrating out' 
(or decoupling) the massive N=1 scalar superfield component of the N=2 vector superfield. 

It is worth mentioning that a positivity of the `discriminant' (under the square root in the 
denominator of eq.~(2.26)) is ensured by a positivity of the BI 
determinant  on the left-hand-side of eq.~(2.17). A causal (no ghosts) 
propagation of the physical fields in the sBI theory is achieved due to the 
Goldstone nature of the whole N=1 vector multiplet and its irreducibility 
with respect to unbroken supersymmetry. The auxiliary field $D$ does not
propagate, with $D=0$ being an on-shell solution to its equation of motion.

The whole non-linear structure of the N=1 sBI action (2.25) is dictated by the
hidden non-linearly realised $S$-supersymmetry whose transformation laws can be
found in ref.~\cite{bg}. It is, therefore, not very surprising that the same 
action (2.25) can be nicely represented as the `non-linear sigma-model'
\cite{bg}
$$ S\low{\rm N=1~GM} = \frac{1}{4}\int d^4x d^2\q \, X + {\rm h.c.}~,
\eqno(2.28)$$
where the chiral $N=1$ superspace lagrangian $X$ obeys a non-linear N=1 
superfield constraint \cite{bg},
$$ X = \frac{1}{4}X\bar{D}^2\bar{X}+W^2~.\eqno(2.29)$$
The uniqueness of the N=1 Goldstone-Maxwell action (2.28) now becomes apparent
because of the identity $X^2=0$. The N=1 chiral superfield $X$ can be
interpreted as the chiral N=1 superfield component in the N=1 superspace 
description of the N=2 vector superfield \cite{rte} (see also subsect.~2.3).

The e.-m. self-duality of the BI action is also naturally generalized to the
N=1 supersymmetric e.-m. self-duality of the N=1 sBI action, when using the N=1
supersymmetric analogue 
$$ S\low{\rm N=1}=  S\low{\rm N=1~GM} +\left[ \frac{i}{2}\int d^4x d^2\q \, 
\tilde{W}^{\a}W_{\a}+ {\rm h.c.}\right] \eqno(2.30)$$
of the bosonic first-order action (2.20). Here the N=1 chiral Lagrange
multiplier superfield $\tilde{W}^{\a}$ has been introduced to enforce the $N=1$
Bianchi identity given by the second equation (2.21) on $W_{\a}$ that is merely
an N=1 chiral superfield in eq.~(2.30). Hence, on the one side, varying 
the action (2.30) with respect to  $\tilde{W}^{\a}$ gives us back the action 
(2.25), whereas, on the other side, varying the action (2.30) with respect
to  $W_{\a}$ instead, solving the arising equation on  $W_{\a}$ in terms of
 $\tilde{W}^{\a}$, and substituting the result back into the action (2.30),
yield {\it the same} sBI action (2.25) in terms of $\tilde{W}^{\a}$. This is
the $N=1$ supersymmetric e.-m. self-duality in terms of N=1 superfiels 
\cite{bg,rte}.

\subsection{N=2 sBI action}

A manifestly N=2 supersymmetric and e.-m. self-dual extension of the 4d BING 
action (1.8) with two real scalars can be constructed in N=2 superspace as the
N=2 generalization of the bosonic BI action (2.1) \cite{k2}. Two massless
Goldstone bosons and Maxwell vector can be unified into a single massless
N=2 vector supermultiplet. The N=2 sBI action \cite{k2} can be considered 
either as the Goldstone action associated with partial breaking of N=4 
supersymmetry down to N=2 in 4d, with the Goldstone-Maxwell N=2 supermultiplet
with respect to unbroken N=2 supersymmetry, or, equivalently, as the 
gauge-fixed N=2 superfield action of a D-3-brane in 
flat six-dimensional ambient spacetime (sect.~1). This action can be most 
easily constructed in the standard N=2 superspace parametrized by the 
coordinates 
$Z^M=(x^{\m},\q^{\a}_i,\bar{\q}^{\dt{\a}i})$ where $\m=0,1,2,3$, $\a=1,2$, 
$i=1,2$, and $\Bar{\q^{\a}_i}=\bar{\q}^{\dt{\a}i}$. The Goldstone-Maxwell N=2 
supermultiplet is described in this N=2 superspace by a {\it restricted} 
chiral (complex scalar) N=2 superfield $W$ \cite{wess,oldrev}. The N=2 
superspace approach automatically implies manifest (linearly realised) N=2 
extended supersymmetry. One cannot, however, use a similar N=4 superspace
approach, in order to construct a 4d sBI/BING action with manifest N=4 
supersymmetry, since a 4d gauge field theory with linearly realised N=4 
supersymmetry merely exists in its on-shell version, in the standard N=4 
superspace in 4d.

The restricted chiral N=2 superfield $W$ is an off-shell irreducible N=2 
superfield satisfying the N=2 superspace constraints
$$ \bar{D}_{\dt{\a}i}W=0~,\qquad D^4W=\bo \bar{W}~,\eqno(2.31)$$
where we have used the following realisation of the supercovariant N=2 
superspace derivatives (with vanishing central charge) \cite{oldrev}:
$$ D^i_{\a}=\fracmm{\pa}{\pa\q^{\a}_i} +i\bar{\q}^{\dt{\a}i}\pa_{\a\dt{\a}}~,
\quad
\bar{D}_{\dt{\a}i}= -\fracmm{\pa}{\pa\bar{\q}^{\dt{\a}i}} 
-i\q^{\a}_i\pa_{\a\dt{\a}}~;\quad D^4 
\equiv \frac{1}{12}D^{i\a}D^j_{\a}D^{\b}_iD_{j\b}~.\eqno(2.32)$$
The first constraint in eq.~(2.31) is just the N=2 generalization of the usual 
N=1 chirality condition, whereas the second one can be considered as the 
generalized {\it reality} condition \cite{sg,oldrev} that has no analogue in 
N=1 superspace. A component solution to eq.~(2.31) in the N=2 chiral 
superspace (parametrized by the coordinates 
$y^{\m}=x^{\m}-\frac{i}{2}\q_i^{\a}\s^{\m}_{\a\dt{\a}}\bar{\q}^{\dt{\a}i}$ and
$\q_{\b}^j$) reads 
$$\eqalign{
W(y,\q)~=~& a(y) + \q^{\a}_i\j^i_{\a}(y)-\ha\q^{\a}_i\vec{\t}\ud{i}{j}
\q^j_{\a}\cdot\vec{D}(y)\cr
~&  +\frac{i}{8}\q_i^{\a}(\s^{\m\n})\du{\a}{\b}\q_{\b}^iF_{\m\n}(y)
-i(\q^3)^{i\a}\pa_{\a\dt{\b}}\bar{\j}_i^{\dt{b}}(y)+\q^4\bo \bar{a}(y)~,\cr}
\eqno(2.33)$$
where we have introduced a complex scalar $a$, a chiral spinor doublet $\j$, 
a real isovector
$\vec{D}=\ha(\vec{\t})\ud{i}{j}D\ud{j}{i}\equiv \ha\tr(\vec{\t}D)$, 
$\tr(\t_m\t_n)=2\d_{mn}$,
and a real antisymmetric tensor $F_{\m\n}$ as the field components of $W$, 
while $F_{\m\n}$ has to satisfy the `Bianchi identity' \cite{oldrev}
$$ \ve^{\m\n\l\r}\pa_{\n}F_{\l\r}=0 ~,\eqno(2.34)$$
whose solution is just given by $F_{\m\n}=\pa_{\m}A_{\n}-\pa_{\n}A_{\m}$ in 
terms of the vector gauge field $A_{\m}$ subject to the gauge transformations 
$\d A_{\m}=\pa_{\m}\l$. The N=2 supersymmetry transformation laws for the 
components can be found e.g., in ref.~\cite{oldrev}.

The well-known N=2 supersymmetric extension of the
Maxwell lagrangian $A=-\frac{1}{4}F^2_{\m\n}$ is given by
$$ \frac{1}{2} \int d^4\q\,W^2=
-a\bo\bar{a}-\frac{i}{2}\j^{\a}_j\pa_{\a\dt{\a}}\bar{\j}^{\dt{\a}j}
-\frac{1}{2}(F^+)^2 +\frac{1}{2}\vec{D}^2~.\eqno(2.35)$$
The Maxwell energy-momentum tensor squared or, equivalently, the EH lagrangian 
$E=\frac{1}{8}(F^+)^2(F^-)^2$, is also easily extended in N=2 superspace,
$$ \int d^4\q d^4\bar{\q}\,W^2\bar{W}^2=(F^+)^2(F^-)^2 
+(\vec{D}^2)^2 -\vec{D}^2F^2+\ldots~.\eqno(2.36)$$
This N=2 supersymmetric generalization of the Euler-Heisenberg lagrangian also
arises as the leading (one-loop) non-holomorphic (non-BPS) contribution to the
N=2 gauge low-energy effective action in the interacting N=2 supersymmetric
quantum field theory of a charged hypermultipet minimally coupled to an N=2 
Maxwell supermultiplet \cite{buch,krev}.  

The gauge-invariant N=2 superfield strength squared, $W^2$, is an 
N=2 chiral but not a restricted N=2 chiral superfield. As is clear from 
eq.~(2.35), the first component of the N=2 anti-chiral superfield  
$K\equiv D^4W^2$ takes the form
$$ K \equiv D^4W^2= -2a\bo\bar{a}-(F^+)^2+\vec{D}^2+\ldots~. \eqno(2.37)$$

It is now straightforward to N=2 supersymmetrize the BI lagrangian (2.11) by 
engineering the proper N=2 superspace invariant, 
$$ L = \frac{1}{2} \int d^4\q\, W^2  +\frac{1}{8}
\int d^4\q d^4\bar{\q}\,\cy(K,\bar{K}) W^2\bar{W}^2~,
\eqno(2.38)$$
whose `formfactor' $\cy(K,\bar{K})$ is dictated by the known bosonic 
structure function $Y(A,E)$ in eq.~(2.18). Note that the 
vector-dependent contributions to the first scalar components of the N=2
superfields $K$ and $\bar{K}$ are simply related to $A$ und $E$ as
$$ K+\bar{K}=4A~,\qquad K\bar{K}=8E~,\eqno(2.39)$$
i.e. they are just the roots of a quadratic equation
$$ k^2-4Ak+8E=0~.\eqno(2.40)$$
We find
$$ \eqalign{
 \cy(K,\bar{K})~=~&
\fracmm{1-\frac{1}{4}(K+\bar{K})-\sqrt{(1-\frac{1}{4}K-\frac{1}{4}\bar{K})^2
-\frac{1}{4}K\bar{K}}}{K\bar{K}} \cr
~=~& 1 + \frac{1}{4}(K+\bar{K})+O(K^2)~.\cr} \eqno(2.41)$$

The proposed N=2 sBI action \cite{k2}
$$\eqalign{
S[W,\bar{W}] ~& = \frac{1}{2} \int d^4x d^4\q\, W^2  + \frac{1}{8}
\int d^4x d^4\q d^4\bar{\q}\,\cy(K,\bar{K}) W^2\bar{W}^2\cr
~& = \frac{1}{2} \int d^4x d^4\q \left\{  W^2 +\frac{1}{4}\bar{D}^4\left[
\cy(K,\bar{K}) W^2\bar{W}^2\right]\right\} \cr
~& = \frac{1}{2} \int d^4x d^4\q \, W^2_{\rm improved}~,\cr}\eqno(2.42)$$
can be nicely rewritten to the `non-linear sigma-model' form
$$ S[W,\bar{W}] = \frac{1}{2} \int d^4x d^4\q\, X ~,\eqno(2.43)$$
where the N=2 chiral superfield $X\equiv W^2_{\rm improved}$ has been 
introduced as a solution to the non-linear N=2 superfield constraint   
$$ X = \frac{1}{4}X\bar{D}^4\bar{X} + W^2~.\eqno(2.44)$$
Eq.~(2.43) is the `improved' non-linear extension of the N=2 Maxwell 
lagrangian in N=2 chiral superspace. The existence of the 
`non-linear sigma-model' form of our action (2.42) implies its uniqueness 
and supports 
its interpretation as the Goldstone action associated with partial breaking 
of N=4 supersymmetry down to N=2, with the N=2 vector multiplet as a Goldstone
multiplet, in a remarkable similarity to the N=1 supersymmetric 
Goldstone-Maxwell theory discussed in the previous subsect.~2.2. 
In particular, eq.~(2.44) can be considered as the N=2 superfield 
generalization of the N=1 superfield non-linear constraint (2.29).

Like the N=1 sBI action, our N=2 action (2.42) does not lead to the 
propagating auxiliary fields $\vec{D}$, despite of the presence of higher 
derivatives to all orders. Though the equations of motion for the auxiliary 
fields do not seem to be algebraic, the kinetic terms for them do not appear,
with $\vec{D}=0$ being an on-shell solution. Non-vanishing expectation values
for fermionic and scalar composite operators in front of the `dangerous' 
interacting terms that could lead to a propagation of the auxiliary fields 
are also forbidden because of the vanishing N=2 central charge and unbroken 
Lorentz- and N=2 super-symmetries. We recall that the N=2 central 
charge $Z$ in abelian N=2 supersymmetric gauge field theories can be 
identified with a (complex constant) vacuum expectation value $\VEV{a}$ of 
the first scalar component of the Maxwell N=2 superfield strength,
$\VEV{W}=\VEV{a}=Z$ (see e.g., ref.~\cite{krev}).

To verify that eq.~(2.42) is the N=2 supersymmetric extension of the N=1 sBI 
action indeed, it is useful to rewrite it in terms of N=1 superfields by 
integrating over a half of the N=2 superspace anticommuting coordinates. 
The standard identification of the N=1 superspace anticommuting 
coordinates,~\footnote{We underline particular values $i=\un{1},\un{2}$ 
of the internal $SU(2)$ indices, and use the N=1 notation \newline ${~~~~~}$ 
$D^2=\frac{1}{2}(D^{\un{1}})^{\a}(D^{\un{1}})_{\a}$ and 
$\bar{D}^2=\frac{1}{2}(\bar{D}_{\un{1}})_{\dt{\a}} 
(\bar{D}_{\un{1}})^{\dt{\a}}$ here.}
$$ \q^{\a}{}_{\un{1}}=
\q^{\a}~,\quad{\rm and}\quad \bar{\q}_{\dt{\a}}{}^{\un{1}}
=\bar{\q}_{\dt{\a}}~, \eqno(2.45)$$
implies the N=1 superfield projection rule
$$ G=\left.G(Z)\right|~,\eqno(2.46)$$
where $|$ means taking a $(\q^{\a}{}\low{\un{2}},
\bar{\q}_{\dt{\a}}{}^{\un{2}})$-independent part
of an N=2 superfield $G(Z)$. As regards the N=2 restricted chiral superfield 
$W$, its N=1 superspace constituents are given by N=1 complex superfields 
$\F$ and $W_{\a}$,
$$ \left.W\right|=\F~, \quad D^{\un{2}}_{\a}\left.W\right|=W_{\a}~,\quad
\frac{1}{2}(D^{\un{2}})^{\a}(D^{\un{2}})_{\a}\left.W\right|=
\bar{D}^2\bar{\F}~,\eqno(2.47)$$
which follow from the N=2 constraints (2.31). The reality condition given by 
the second equation (2.31) implies the N=1 superfield Bianchi identity
$$ D^{\a}W_{\a}=\bar{D}_{\dt{\a}}\bar{W}^{\dt{\a}}~,\eqno(2.48)$$
as well as the relations
$$\eqalign{
 \left.K\right|~=~ & D^2\left( W^{\a}W_{\a} + 2\F\bar{D}^2\bar{\F}\right)~.\cr
(\bar{D}_{\un{2}})^{\dt{\a}}\left.K\right|~=~ & 
2iD^2\pa^{\dt{\a}\b}\left(W_{\b}\F\right)~,
\cr
(\bar{D}_{\un{2}})_{\dt{\a}}(\bar{D}_{\un{2}})^{\dt{\a}}\left.K\right|~=~ &
-4D^2\pa_{\m}\left( \F\pa^{\m}\F\right)~,\cr}\eqno(2.49)$$
together with their conjugates. Eqs.~(2.47), (2.48) and (2.49) are enough to 
perform a reduction of any N=2 superspace action depending 
upon $W$ and $\bar{W}$ into N=1 superspace by differentiation, 
$$\eqalign{
\int d^4\q ~\to~& \int d^2\q\, \frac{1}{2} (D^{\un{2}})^{\a}(D^{\un{2}})_{\a}~,
\cr
\int d^4\q d^4\bar{\q} ~\to~& \int d^2\q d^2\bar{\q}\, 
\frac{1}{2} (D^{\un{2}})^{\a}
(D^{\un{2}})_{\a}\frac{1}{2} 
(\bar{D}_{\un{2}})_{\dt{\a}}(\bar{D}_{\un{2}})^{\dt{\a}}~.\cr}
\eqno(2.50)$$
It is now straightforward to calculate the N=1 superfield form of the N=2 
action (2.42). For our purposes, it is enough to notice that the first term in 
eq.~(2.42) gives rise to the kinetic terms for the N=1 chiral superfields
 $\F$ and $W_{\a}$,
$$ {\rm Re}\int d^2\q\,(\frac{1}{2}W^{\a}W_{\a} 
+\F\bar{D}^2\bar{\F})~,\eqno(2.51)$$
whereas the N=1 vector multiplet contribution arising from the second term in 
eq.~(2.42) is given by
$$\frac{1}{8} 
\int d^2\q d^2\bar{\q}\,\cy(\left.K\right|,\bar{K}\left.\right|)W^{\a}W_{\a}
\bar{W}_{\dt{\a}}\bar{W}^{\dt{\a}} +\ldots~,\eqno(2.52)$$
where the dots stand for $\F$-dependent terms. The $W$-dependent contributions
 of eqs.~(2.51) and (2.52) exactly coincide with the N=1 supersymmetric 
extension (2.25) of the BI action  after taking into account that the vector 
field dependence in the first component of the N=1 superfield $\left.K\right|$
is given by
$$ \left.K\right| =\left. D^4W^2\right|= 
\left. 2D^2 ( \frac{1}{2} W^{\a}W_{\a}+\F\bar{D}^2\bar{\F})\right|
= -(F^+)^2+D^2+\ldots, \eqno(2.53)$$
and similarly for  $\bar{K}\left.\right|$.

The dependence of the N=2 sBI action upon the N=1 chiral part $\F$ of the N=2 
vector multiplet is clearly of most interest, since it is entirely dictated by
N=2 extended supersymmetry and electric-magnetic self-duality. Let's now take 
$W_{\a}=0$ in the action (2.42), and  calculate merely the leading terms 
depending upon $\F$ and $\bar{\F}$ there. After some algebra one gets 
the following N=1 superspace action:
$$ S[\F,\bar{\F}]=\int d^4x d^2\q d^2\bar{\q}\left[ \F\bar{\F} 
-4(\F\pa^{\m}\F)(\bar{\F}\pa_{\m}\bar{\F})
+4\pa^{\m}(\F\bar{\F})\pa_{\m}(\F\bar{\F})\right]
+\ldots~,\eqno(2.54)$$
where the dots stand for the higher order terms depending upon the derivatives
of $\cy$. The field components of the N=1 chiral superfield $\F$ are 
conveniently defined by the projections
$$ \left.\F\right|=\fracmm{1}{\sqrt{2}}\f\equiv\fracmm{1}{\sqrt{2}}(P+iQ)~,
\quad D_{\a}\left.\F\right|=\j_{\a}~,\quad D^2\left.\F\right|=F~,\eqno(2.55)$$
where $P$ is a real physical scalar, $Q$ is a real physical pseudo-scalar, 
$\j_{\a}$ is a chiral physical spinor, and $F$ is a complex auxiliary field. 
It is not difficult to check that the kinetic terms for the auxiliary 
field components $F$ and $\bar{F}$ cancel in eq.~(2.54), as they should. This 
allows us to simplify a calculation of the quartic term in eq.~(2.54) even 
further by going on-shell, i.e. assuming that $\bo\f=F=0$ there, even though 
it is not really necessary. A simple calculation now yields 
$$ S[ \f,\bar{\f} ]=\int d^4x\, \left\{ \pa^{\m}\f\pa_{\m}\bar{\f} -
2(\pa_{(\m}\f\pa_{\n)}\bar{\f})^2 + (\pa_{\m}\f\pa^{\m}\bar{\f})^2\right\}~,
\eqno(2.56)$$
which exactly coincides with the leading terms in the derivative expansion of 
the Nambu-Goto (NG) action
$$ S=-\int d^4x\, \sqrt{-\det(\h_{\m\n}+\pa_{\m}P\pa_{\n}P
+\pa_{\m}Q\pa_{\n}Q)}~.\eqno(2.57)$$
Eq.~(2.57) yields the effective action of a (static-gauge) 3-brane in flat
six-dimensional ambient spacetime, with the Goldstone scalars $(P,Q)$ being 
two collective coordinates corresponding to a `transverse' motion of the 
3-brane.~\footnote{An N=1 superspace description of the two transverse 3-brane
coordinates in terms of N=1 chiral, \newline ${~~~~~}$ complex linear and real
linear Goldstone superfields was recently obtained in ref.~\cite{stony}.} 
A 3-brane solution to $(1,0)$, 6d super-Maxwell theory coupled to 
chiral (scalar) multiplets was constructed in ref.~\cite{hlp}. 
The solution of ref.~\cite{hlp} breaks translational invariance in two spacial
directions and half of the 6d supersymmetry. This observation strongly 
indicates on a possible six-dimensional origin of our four-dimensional N=2 
supersymmetric BI action that should be derivable by dimensional reduction 
from a supersymmetric BI action in six spacetime dimensions after identifying 
the extra two components of a six-dimensional abelian vector potential with 
the scalars $P$ and $Q$. The very existence of the super-BI action in six 
dimensions is enough to ensure the Goldstone nature of scalars in eq.~(2.56) 
and (2.57), as well as the {\it off-shell} invariance of our 4d, N=2 action under 
constant shifts of these scalars. Finding this 6d, manifestly (1,0)
supersymmetric BI action, which can be considered as the top or {\it master} 
sBI action in superspace, is one of our main results in this paper (see sect.~3).

To indicate on the possibility of adding some additional structure given by
a magnetic FI term into our 4d, N=2 theory, it is worth mentioning here that 
the N=2 superspace constraints (2.31) imply
\def\bo{{\raise-.3ex\hbox{\large$\Box$}}}
$$\bo\left( D^{ij}W-\bar{D}^{ij}\bar{W}\right)=0~.\eqno(2.58)$$
This means that the function ${\rm Im}\,(D^{ij}W)$ is harmonic, and, 
therefore, it should be constant,~\footnote{We assume 
that all components of the superfield $W$ are regular in spacetime.} i.e.
$$ D^{ij}W-\bar{D}^{ij}\bar{W}=4iM^{ij}~.\eqno(2.59)$$
Taking into account a constant (FI) vector $\vec{M}$ in the constraint (2.59) is equivalent to adding 
a `{\it magnetic\/}' Fayet-Iliopoulos (FI) term to the dual action \cite{apt}. 
The FI term can be formally removed from the constraint (2.59) by a field redefinition of 
$W$, i.e. at the expense of adding a constant imaginary part to the auxiliary 
scalar triplet $\vec{D}$ of the N=2 vector multiplet in eq.~(2.33). Let's recall that 
the APT model \cite{apt} is defined by adding the usual (electric) and magnetic FI terms to the 
general (Seiberg-Witten-type) N=2 chiral action in terms of $W$ \cite{apt,oldrev}.  

The 4d, N=2 `Bianchi identity' can be enforced by introducing an unconstrained 
real N=2 superfield Lagrange multiplier (known as the {\it Mezincescu} pre-potential
\cite{mez}) $\vec{L}=\ha(\vec{\t})\ud{i}{j}L\ud{j}{i}\equiv \ha\tr(\vec{\t}L)$
in the first-order N=2 superspace action ({\it cf.} ref.~\cite{iz})
$$\eqalign{
 S[W,\bar{W}] \to S[W,\bar{W};L] ~=~& S[W,\bar{W}]
+ i\int d^4x d^4\q d^4\bar{\q}\, L_{ij}
\left(D^{ij}W-\bar{D}^{ij}\bar{W}\right)~,\cr
~=~&  S[W,\bar{W}] +\left[ i\int d^4x d^4\q\, WW_{\rm magn.} 
+{\rm h.c.}\right]~,\cr}\eqno(2.60)$$
where the N=2 superfield $W$ is now a chiral (unrestricted) N=2 superfield, 
while
$$ W_{\rm magn.} \equiv \bar{D}^4D^{ij}L_{ij} \eqno(2.61)$$
is the dual or `magnetic' N=2 superfield strength that automatically 
satisfies the N=2 constraints (2.31) due to its defining equation (2.61). 
Varying the action (2.60) with respect to $W$, solving the resulting algebraic
equation on $W$ in terms of $W_{\rm magn.}$, and substituting the result back
into the action (2.60), results in the dual N=2 action 
$S[W_{\rm magn.},\bar{W}_{\rm magn.}]$ that
takes exactly {\it the same} form as eq.~(2.42). In other words, it is 
self-dual with respect to the N=2 supersymmetric electric-magnetic duality. 
Of course, the (1,0) supersymmetric BI action in six spacetime dimensions 
(sect.~3) cannot be e.-m. self-dual since the dual to a vector is again going to be 
a vector in four spacetime dimensions only.

The 4d, N=2 Maxwell multiplet considered in this subsection can be interpreted 
as the Goldstone multiplet associated with partial spontaneous breaking of 
rigid N=4 supersymmetry down to N=2 supersymmetry, with the action (2.42) 
being the corresponding Goldstone-Maxwell N=2 superfield action ({\it cf.}
ref.~\cite{bik}). The whole non-linear structure of this action dictated by the
non-linear superfield constraint (2.44) should therefore be entirely 
determined by hidden (non-linearly realised, or broken) N=2 supersymetry 
({\it cf.} refs.~\cite{aps,bg}). The transformation laws of the spontaneously 
broken N=2 supersymmetry can be deduced \cite{pro} from the general theory of 
non-linear realizations \cite{bagger,cwz}, despite of the fact that the general theory 
\cite{bagger,cwz} does not give us any clues about a non-perturbative construction of the 
Goldstone actions beyond the Noether (trials and errors, order-by-order) method. 

The invariance of our action (2.42) under constant shifts 
of the N=2 superfield strength $W$ is, of course, a necessary condition for its
Goldstone interpretation. It is easy to verify this symmetry if $W$ is subject
to the on-shell condition $\bo W=0$. The 6d super-BI action (sect.~3) 
dimensionally reduced down to four dimensions automatically implies this 
symmetry off-shell.

\section{Gauge-fixed `spacetime-filling' D-5-brane action}

In this section we generalize the results of sect.~2 by constructing a (1,0) 
supersymmetric Born-Infeld-Goldstone action in 6d superspace.

\subsection{Group theory: $SU(4)$ versus $SO(1,5)$}

Let's now consider flat six-dimensional (6d) spacetime with a Minkowski 
signature $\h_{\m\n}={\rm diag}(+,-,-,-,-,-)$, where the vector indices take
values $\m,\n=0,1,2,3,4,5$. The 6d 
Lorentz group $SO(1,5)$ has rank three, so that there are three independent 
Casimir eigenvalues in 6d instead of two in the 4d case (sect.~2). The obvious 
choice of the independent Lorentz-invariant products of a 6d abelian gauge 
field strength $F_{\m\n}=\pa_{\m}A_{\n}-\pa_{\n}A_{\m}$ is given by
$$ \eqalign{
F^2 ~&~ \equiv F_{\m\n}F^{\m\n}~,\cr
F^4 ~&~ \equiv F_{\m\n}F^{\n\l}F_{\l\r}F^{\r\m}~,\cr
F^6 ~&~ \equiv F_{\m\n}F^{\n\l}F_{\l\r}F^{\r\s}F_{\s\t}F^{\t\m}~.\cr}
\eqno(3.1)$$
It is straightforward to calculate any Lorentz-invariant function of 
$F_{\m\n}$ in terms of the invariants (3.1). In the BI case, we find
$$\eqalign{
-\det\left( \h_{\m\n}+F_{\m\n} \right)~ & = ~ 
1+ \fracmm{1}{2}F^2+\fracmm{1}{2^3}(F^2)^2 -\fracmm{1}{2^2}F^4 \cr
~& ~ + \fracmm{1}{3\cdot 2^4}(F^2)^3 - \fracmm{1}{2^3}F^2F^4
+\fracmm{1}{2\cdot 3}F^6~,
\cr}\eqno(3.2)$$ 
and 
$$\eqalign{
-\,\sqrt{-\det\left(\h_{\m\n}+F_{\m\n}\right)} & = 
-1 - \fracmm{1}{2^2}F^2
 - \fracmm{1}{2^3}\left[ \fracmm{1}{4}(F^2)^2-F^4\right] \cr
& +\fracmm{1}{2^5}\left[ F^2F^4 -\fracmm{1}{2^2\cdot 3}(F^2)^3 -
\fracmm{2^3}{3}F^6\right]\cr
& -\fracmm{1}{2^8}\left[ 3^2(F^2)^2F^4-\fracmm{2^4}{3}F^2F^6 -2(F^4)^2
-\fracmm{7}{2^3\cdot 3}(F^2)^4\right] +O(F^{10}),\cr}\eqno(3.3)$$
where we have used the expansion
$$ \sqrt{1+x}=1+\fracmm{1}{2}x-\fracmm{1}{2^3}x^2+\fracmm{1}{2^4}x^3-
\fracmm{5}{2^7}x^4+O(x^5)~.\eqno(3.4)$$

The leading term in the expansion (3.3) of the BI action is given by the
6d Maxwell lagrangian, $-\frac{1}{4}F^2$, as expected. The quartic (of the 
fourth order in spacetime derivatives) terms in eq.~(3.3) occur in the same 
combination as in the 4d case (see the right-hand-side of the first line of 
eq.~(2.11) in subsect.~2.1), $\frac{1}{8}\left[\frac{1}{4}(F^2)^2-F^4\right]$, 
equal to the 6d Maxwell energy-momentum tensor squared. It agrees with 
(i) earlier perturbative calculations of the gauge low-energy 
effective action of open superstrings \cite{gw}, and (ii) restrictions implied 
by supersymmetry in 10d \cite{gv} and 6d \cite{brs}. In order to make these
restrictions {\it manifest} in 6d, it is useful to switch to a four-component 
(spinor) $SU(4)$ notation in 6d, which is similar to the two-component 
spinor notation in 4d, by using a complex `rotation' of Lie algebra of 
$SO(1,5)$ to that of $SU(4)$ \cite{hst}. 

Let $\G^{m}$ be $8\times 8$ gamma matrices in 6d, that satisfy a 
Clifford algebra~\footnote{In order to distinguish between 6d vector and
spinor indices both denoted by greek letters, we  \newline ${~~~~~}$ also 
introduce here latin letters to denote tangent space vector components. 
They are trivially \newline ${~~~~~}$ related to spacetime vector 
components, $V^{\m}=e^{\m}_m V^m$, via a flat 6-bein $e^{\m}_m=\d^{\m}_m$. 
This notation \newline ${~~~~~}$ shall, however, be abandoned in the
next subsections where all vector indices, if any, are hidden, 
\newline ${~~~~~}$ while latin indices are used to denote 
internal $SU(2)$ symmetry.}
$$ \{ \G^{m}, \G^{n} \}=2\h^{mn}~.\eqno(3.5)$$
Let's choose a representation of these matrices in the form 
$$  \G^{m}= \left( \begin{array}{cc} 0 & (\S^{m})_{\a\dt{\b}} \\
(\tilde{\S}^{m})^{\dt{\a}\b} & 0 \end{array} \right)~,\qquad
\a=1,2,3,4~,\eqno(3.6)$$
where the $4\times 4$ matrices $\S^{m}$ and $\tilde{\S}^{m}$ have been 
introduced. They have to satisfy the relations
$$ \eqalign{
 \S^{m}\tilde{\S}^{n}+ \S^{n}\tilde{\S}^{m}=2\h^{mn}~,\cr 
 \tilde{\S}^{m}\S^{n}+ \tilde{\S}^{n}\S^{m}=2\h^{mn}~.\cr} 
\eqno(3.7)$$
A solution to eq.~(3.7) exists in the form
$$ \S^{m}=({\bf 1},\g^i)~,\quad
\tilde{\S}^{m}=({\bf 1},-\g^i)~,\qquad m=(0,i)~,\quad  i=1,2,3,4,5~,
\eqno(3.8)$$
where $\g^i$ are hermitian $4\times 4$ gamma matrices in {\it five} euclidean 
dimensions,
$$ \{ \g^i,\g^j\}=2\d^{ij}~.\eqno(3.9)$$
An explicit representation of $\g^i=(\vec{\g},\g_4,\g_5)$, with 
$\vec{\g}$ standing for $(\g_1,\g_2,\g_3)$ and $\g_5=\g_1\g_2\g_3\g_4$, 
is given by \cite{van} 
$$\vec{\g}=\left(  \begin{array}{cc} 0 & -i\vec{\s} \\
i\vec{\s} & 0 \end{array} \right)~,\qquad
\g_4=\left(  \begin{array}{cc} I_2 & 0 \\ 0 & -I_2 \end{array}\right)~,
\qquad \g_5=\left(  \begin{array}{cc}  0 & -I_2 \\  -I_2 & 0 \end{array} 
\right)~,\eqno(3.10)$$
where $\vec{\s}$ are Pauli matrices, $\s_1\s_2\s_3=iI_2$.

The 6d Lorentz generators in a non-chiral spinor representation are given by
$$  \G^{mn}=\frac{1}{4}\[ \G^m,\G^n \]=\left(
\begin{array}{cc} (\S^{mn})\du{\a}{\b} & 0 \\ 0 & 
(\tilde{\S}^{mn})\ud{\dt{\a}}{\dt{\b}} \end{array} \right)~,
\eqno(3.11)$$ 
where 
$$ \S^{mn}=\frac{1}{4}\left( \S^m\tilde{\S}^n-\S^n\tilde{\S}^m\right)~,
\eqno(3.12)$$
and similarly for $\tilde{\S}^{mn}$. The chiral spinor generators
$\S^{mn}$ satisfy the Lorentz algebra
$$ \[ \S^{mn},\S^{pq} \] = \h^{mq}\S^{np} +\h^{nq}\S^{pm} +\h^{mp}\S^{qn}
+\h^{np}\S^{mq}~,\eqno(3.13)$$
while they can be used to convert any antisymmetric tensor $F_{mn}$ into 
a traceless $4\times 4$ matrix $\hat{F}\du{\a}{\b}$ as follows: 
$$\hat{F}\du{\a}{\b}=(\S^{mn})\du{\a}{\b}F_{mn}~,\quad \tr\,\hat{F}=0~.
\eqno(3.14)$$ 

As a preparation for (1,0) supersymmetrization of 6d BI action 
(subsect.~3.3),
let's introduce another set of independent Lorentz-invariant $F$-products:
$$
\tr\,\hat{F}^2~,\quad \tr\,\hat{F}^4~,\quad \tr\,\hat{F}^6~.\eqno(3.15)$$
They are, of course, linearly related to those of eq.~(3.1). We find
$$ \eqalign{
\tr\,\hat{F}^2 = & ~-2F^2~,\cr
\tr\,\hat{F}^4 = & ~3(F^2)^2 -4F^4~,\cr
\tr\,\hat{F}^6 = & ~-32F^6 - \fracmm{15}{2}F^2\left[ (F^2)^2-4F^4\right]~.\cr}
\eqno(3.16)$$
It is straightforward to verify that e.g., $\tr\,\hat{F}^8$ is not 
independent, being a function of those in eq.~(3.16), namely,
$$ \tr\,\hat{F}^8 = \frac{1}{4}(\tr\,\hat{F}^4)^2 +
\frac{4}{3}\tr\,\hat{F}^6\tr\,\hat{F}^2 -\frac{3}{4}\tr\,\hat{F}^4
(\tr\,\hat{F}^2)^2 +\frac{5}{48}(\tr\,\hat{F}^2)^4~.\eqno(3.17)$$ 

After some algebra, we find
$$\eqalign{
-\det\left( \h_{mn}+F_{mn} \right)~ & = ~ 
1 - \fracmm{1}{2^2}\tr\,\hat{F}^2 +\fracmm{1}{2^4}\left[
\tr\,\hat{F}^4 -\fracmm{1}{4}(\tr\,\hat{F}^2)^2\right] \cr
~& ~~ +\fracmm{1}{2^8}\left[\tr\,\hat{F}^2\tr\,\hat{F}^4 -\fracmm{1}{6}
(\tr\,\hat{F}^2)^3 -\fracmm{4}{3}\tr\,\hat{F}^6\right]~,\cr}\eqno(3.18)$$
and
$$\eqalign{
\sqrt{ -\det\left( \h_{mn}+F_{mn} \right)} ~ & = ~
1-\fracmm{1}{2^3}\tr\,\hat{F}^2 +\fracmm{1}{2^5}\left[
\tr\,\hat{F}^4 -\fracmm{1}{2}(\tr\,\hat{F}^2)^2\right] \cr
~& ~~ -\fracmm{1}{3\cdot 2^7}\tr\,\hat{F}^6 +
\fracmm{3}{2^9}\tr\,\hat{F}^2\tr\,\hat{F}^4 -
\fracmm{7}{3\cdot 2^{10}}(\tr\,\hat{F}^2)^3 \cr
~ & ~~ -\fracmm{1}{2^{11}}(\tr\,\hat{F}^4)^2 +
\fracmm{5}{2^{12}}\tr\,\hat{F}^4(\tr\,\hat{F}^2)^2
-\fracmm{1}{3\cdot 2^{10}}\tr\,\hat{F}^6\tr\,\hat{F}^2 \cr
~& ~~ - \fracmm{3\cdot 5}{2^{15}}(\tr\,\hat{F}^2)^4 +O(F^{10})~.\cr}
\eqno(3.19)$$

The easiest way to get the key equations (3.2) and (3.16) is to take advantage
of their 6d Lorentz invariance by choosing $F_{mn}$ in the form
$$ F_{mn} = \left( \begin{array}{cccccc}  & \l_1 & & & & \\
-\l_1 & & & & & \\ & & & \l_2 & & \\  & & -\l_2 & & & \\ 
 & & & & & \l_3  \\   & & & & -\l_3 & \end{array} \right) 
\eqno(3.20)$$ 
similar to that of eq.~(2.2), in terms of three independent real 
`eigenvalues' $\vec{\l}$. Eq.~(3.2) then amounts to a classical Miura 
transform in terms of symmetric polynomials of $\vec{\l}$, 
whereas the linear transform (3.16)
becomes apparent when using a basis comprising all independent antisymmetric 
products of $\g$-matrices. The coefficients are, of course, Lorentz-invariant
and independent upon the representation of $\g$-matrices used to calculate 
them. Similar techniques were used for a calculation of perturbative anomalies
in chiral 6d supersymmetric gauge field theories and 6d supergravity 
\cite{aw,kano}. 

\subsection{6d Maxwell (1,0) supermultiplet in superspace}

The (1,0) superspace techniques in 6d were proposed e.g., in 
refs.~\cite{nils,hst}. In this subsection we briefly review the 
construction of ref.~\cite{hst}, and then extend it for a later use 
in the next subsect.~3.3.   

Chiral 6d spinors can be equivalently represented by
{\it symplectic} Majorana-Weyl (MW) spinors $\j^{\a}_i$ carrying an extra 
$SU(2)\cong Sp(1)$ index $i=1,2$ and obeying the MW-type condition \cite{kto} 
$$ (\j^{\a}_i)^*\equiv \bar{\j}^{\dt{\a}i}=\ve^{ij}B\ud{\dt{\a}}{\b}\j^{\b}_j
~,\eqno(3.21)$$
where $\ve^{ij}$ is antisymmetric Levi-Civita symbol, $\ve^{ij}\ve_{kj}=
\d^i_k$, and the matrix $B$ can be chosen to be unitary and antisymmetric,
$ BB^{\dg}=1$ and $B^T =-B$. The existence of the matrix $B$ follows from the
uniqueness of a non-trivial representation of 6d Clifford algebra (3.5). 
Using symplectic MW spinors in 6d allows one to directly compare 6d chiral 
supersymmetry to N=2 extended supersymmetry in 4d (subsect.~2.3) where 
symplectic MW spinors also naturally appear and make the internal $SU(2)$
symmetry manifest \cite{kty}.

The matrix $B$ relates dotted and undotted 6d spinor indices, while any 6d
vector index ($\un{6}$ of $SO(1,5)$) can be traded for a pair of undotted 
spinor indices ($\un{6}$ of $SU(4)$) by using a matrix 
$$ (\S^m)_{\a\dt{\b}}B\ud{\dt{\b}}{\b}\equiv (\S^m)\low{\a\b}=
-(\S^m)\low{\b\a}~. \eqno(3.22)$$
We use here the same notation as in ref.~\cite{hst}. The $SU(2)$ indices are
raised and lowered according to the `North-West/South-East' rule,
$$ V^i=\ve^{ij}V_j~,\qquad V_i=V^j\ve_{ji}~,\eqno(3.23)$$
whereas antisymmetric pairs of spinor indices are raised and lowered by using
totally antisymmetric Levi-Civita symbols,
$$ V^{\a\b}=\frac{1}{2}\ve^{\a\b\g\d}V_{\g\d}~,\qquad
  V_{\a\b}=\frac{1}{2}\ve_{\a\b\g\d}V^{\g\d}~.\eqno(3.24)$$
Note the identities \cite{hst}
$$ (\S^m)_{\a\b} (\S^n)^{\b\a}=4\h^{mn}~,\qquad 
 (\S^m)_{\a\b} (\S_m)_{\g\d}=-2\ve_{\a\b\g\d}~.\eqno(3.25)$$

The supercovariant derivatives $D^i_{\a}$ in flat 6d, (1,0) superspace 
$Z^A=(x^m,\q^{\a}_i)$, with the Grassmann anticommuting coordinates 
$\q^{\a}_i$ being a symplectic MW spinor, satisfy a (1,0) supersymmetry 
algebra
$$ \{ D^i_{\a},D^j_{\b}\} =i\ve^{ij}\pa_{\a\b}~.\eqno(3.26)$$ 

It is clear from eq.~(3.26) that imposing the $SU(2)$-covariant chirality
condition, $D^i_{\a}\F=0$, on a 6d superfield $\F$ implies $\pa_{\a\b}\F=0$,
i.e. $\F=const$.  Introducing 6d chiral superfields is, nevertheless, possible
at the expense of breaking the $SU(2)$ symmetry \cite{nils}. 

A massless (1,0) vector multiplet in 6d superspace is described by a
symplectic MW spinor superfield strength  $W^{\a}_i$ that satisfies some
additional off-shell constraints. The superspace constraints in supersymmetric 
gauge field theories are usually imposed on the invariant field strengths 
$F_{AB}$ defined by an algebra of the gauge- and super-covariant superspace 
derivatives $\cd_A=D_A+iA_A$ \cite{bw},
$$ \[ \cd_A,\cd_B\} =t\du{AB}{C}\cd_C + iF_{AB}~,\eqno(3.27)$$
where $t\du{AB}{C}$ is the flat 6d superspace torsion tensor. In particular,
one has
$$ \{  \cd^i_{\a},\cd_{\b}^j \} = i\ve^{ij}\cd_{\a\b} + i F^{ij}_{\a\b}~.
\eqno(3.28)$$
The off-shell (1,0) vector supermultiplet constraints read \cite{hst}
$$F^{ij}_{\a\b}=0~.\eqno(3.29)$$
It follows from eq.~(3.29) and the Bianchi identities 
$D_{[A}F_{BC\}}=0$ related to the defining eq.~(3.27) that 
$$ F_{m\a i}=(\S_m)_{\a\b}W^{\b}_i~,\eqno(3.30)$$
where $W^{\b}_i$ is further constrained by 
$$ D^i_{\a}W^{\b}_j = \hat{F}\du{\a}{\b}\d\ud{i}{j} 
+ \d\du{\a}{\b}Y\ud{i}{j}~,\quad {\rm while} \quad 
\tr\hat{F}=Y\ud{i}{i}=0~.\eqno(3.31)$$
The leading superfield component of $\hat{F}\du{\a}{\b}$ can be identified 
with the 6d Maxwell field strength (3.14), whereas $Y\ud{i}{j}$ is just the 
$SU(2)$ triplet of the scalar auxiliary fields. The fermionic superpartners 
are given by $\left.W^{\a}_i\right|=\j^{\a}_i$.

The 6d superspace constraints on $W^{\a}_i$ can be solved \cite{hst} in terms
of an unconstrained superfield $V_{ij}$ that is the 6d analogue of the 4d, N=2
Mezincescu prepotential in eqs.~(2.60) and (2.61). For example, in a WZ gauge,
one finds~\cite{bik}
$$ V_{ij}(x,\q) = \q^{\a}_i \q^{\b}_jA_{[\a\b]}(x) +
\q^{\a}_i\q^{\b}_j\q^{\g}_k\ve_{\a\b\g\d}\j^{\d k}(x) 
+ (\q^4)_{ijkl}Y^{kl}(x)~,\eqno(3.32)$$
where $A_{[\a\b]}=(\S^m)_{\a\b}A_m$ is the abelian vector gauge field. It is,
therefore, possible to quantize the (1,0) vector multiplet directly in 6d 
superspace. When being interpreted as the 6d Goldstone-Maxwell multiplet, it
should be associated with partial spontaneous breaking of 6d extended chiral 
(2,0) supersymmetry down to (1,0) supersymmetry since the Goldstone fermion 
$\j^{\a}_i(x)$ and the (1,0) anticommuting spinor coordinate $\q^{\a}_i$ in
eq.~(3.32) have the same chirality \cite{bik}. 

The fermionic superfield $W^{\a}_i$ transforms in a representation 
$\un{4}\times\un{2}$ under the symmetry group $SU(4)\times SU(2)$. 
It is convenient to 
describe an irreducible product of $W$'s in terms of a Young tableau 
({\it cf.} refs.~\cite{hst,oldrev}).~\footnote{The same applies to the 
products of anticommuting superspace coordinates $\q^{\a}_i$.} We shall use

\setlength{\unitlength}{0.2mm}

$$ \eqalign{
W^{\a}_i & ~\sim~~ \youngS ~\sim~  \un{4}\times\un{2}~,\cr
W^{\a\b}_{ij} & ~\sim~~ \youngT ~\sim~ \un{6}\times\un{3}~,\cr
W_{ijkl} &  ~\sim~~ \youngX ~\sim~ \un{1}\times\un{5}~,\cr
(W^6)^{ij}_{\a\b} & ~\sim~~ \youngY ~\sim~ \un{6}\times\un{3}~,\cr
(W^8) & ~\sim~~ \youngZ ~\sim~ \un{1}\times\un{1}~,\cr}
\eqno(3.33)$$
where each Young tableau describes an $SU(4)$ irreducible representation 
(irrep). The corresponding $SU(2)$ irrep is obtained by a reflection of the
 Young tableau about the main diagonal \cite{hst}. In their explicit form,
the products defined by eq.~(3.33) are given by 
$$ \eqalign{
W^{\a\b}_{ij} ~=~~ & W^{[\a}_{(i}W^{\b]}_{j)}~,\cr
W_{ijkl} ~=~~ & \ve_{\a\b\g\d}W^{\a}_{(i}W^{\b}_jW^{\g}_kW^{\d}_{l)}~~, \cr
(W^6)^{ij}_{\a\b} ~=~~ & \fracmm{\bvec{\pa}}{\pa W^{[\a}_{(i}}
\fracmm{\bvec{\pa}}{\pa W^{\b]}_{j)}}(W^8)~~,\cr
(W^8) ~=~~ & \prod_{\a,i}W^{\a}_i~~,\cr} \eqno(3.34)$$
where all (anti)symmetrizations are defined with unit weight. 
Note the identities
$$ W^{(\a}_iW^{\b}_jW^{\g)}_k=0~,\qquad (W^9)=0~.\eqno(3.35)$$

A similar description applies to the irreducible products of the supercovariant 
derivatives in flat 6d superspace. We find ({\it cf.} refs.~\cite{hst,oldrev})
$$ \eqalign{
D^i_{\a}  & ~\sim~~ \youngA ~\sim~  \un{\bar{4}}\times\un{2}~,\cr
D_{(\a\b)} & ~\sim~~ \youngB ~\sim~  \un{\Bar{10}}\times\un{1}~,\cr  
D^{ij}_{[\a\b]} & ~\sim~~ \youngC ~\sim~  \un{6}\times\un{3}~,\cr
(D^3)^{\a}_{ijk} & ~\sim~~ \youngH ~\sim~  \un{4}\times\un{4}~,\cr
(D^3)^i_{\a\b\g} & ~\sim~~ \youngK ~\sim~  \un{20}\times\un{2}~,\cr
D^{ijkl} & ~\sim~~ \youngD ~\sim~  \un{1}\times\un{5}~,\cr
(D^4)^{ij}_{\a\b\g\d} & ~\sim~~ \youngJ ~\sim~  \un{15}\times\un{3}~,\cr
(D^4)_{\a\b\g\d} & ~\sim~~ \youngI ~\sim~  \un{20'}\times\un{1}~,\cr
(D^5)^{i}_{\a\b\g} & ~\sim~~ \youngL ~\sim~  \un{\Bar{20}}\times\un{2}~,\cr
(D^5)^{ijk}_{\a} & ~\sim~~ \youngM ~\sim~  \un{\bar{4}}\times\un{4}~,\cr
(D^6)^{\a\b} & ~\sim~~ \youngR ~\sim~  \un{10}\times\un{1}~,\cr
(D^6)_{\a\b}^{(ij)} & ~\sim~~ \youngE ~\sim~  \un{6}\times\un{3}~,\cr
(D^7)^{\a}_i & ~\sim~~ \youngP ~\sim~  \un{4}\times\un{2}~,\cr
(D^8) & ~\sim~~ \youngF ~\sim~  \un{1}\times\un{1}~,\cr}
\eqno(3.36)$$
where we have used the boxes with dots, as in ref.~\cite{hst}, in order to
distinguish these Young tableaux from those of eq.~(3.33). Yet another reason 
is the fact that the fundamental representation $\un{4}$ of $SU(4)$ is complex,
so that the positioning of the $\a$-index in $W^{\a}_i$ and $D_{\a}^i$ matters.
The irreps $\un{10}$ and  $\un{20}$ of $SU(4)$ are also complex, whereas the
irreps $\un{6}$, $\un{15}$ and $\un{20'}$ are real.

Since the supercovariant derivatives do not just anticommute but
satisfy the algebra (3.26), there are ambiguities in defining their products
according to eq.~(3.36). For instance, eq.~(3.26) implies the identity
\cite{hst}  
$$ D_{\a}^iD_{\b}^j =\frac{i}{2}\ve^{ij}\pa_{\a\b} + \ve^{ij}D_{\a\b}
+ D^{ij}_{\a\b}~,\eqno(3.37)$$
where 
$$ D_{\a\b}=\frac{1}{2}D_{i(\a}D^i_{\b)}~,\quad 
D^{ij}_{\a\b}= D_{[\a}^{(i}D_{\b]}^{j)}~.\eqno(3.38)$$
In other words, when the products of $D$'s are defined in the same way as
those of $W$'s, all their ambiguities are just total derivatives in spacetime.
Hence, we can use the same explicit definitions as in eq.~(3.34), {\it viz.}
$$ \eqalign{
D^{ijkl} ~=~~ & \ve^{\a\b\g\d}D_{\a}^{(i}D_{\b}^jD_{\g}^kD_{\d}^{l)}~~, \cr
(D^6)^{\a\b}_{ij} ~=~~ & \fracmm{\bvec{\pa}}{\pa D_{[\a}^{(i}}
\fracmm{\bvec{\pa}}{\pa D_{\b]}^{j)}}(D^8)~~,\cr
(D^8) ~=~~ & \prod_{\a,i}D_{\a}^i~~,\cr} \eqno(3.39)$$
as long as they are going to be integrated over all spacetime coordinates. It
is just the case in all the equations to be introduced in the next 
subsect.~3.3.~\footnote{The products defined by eq.~(3.38) and the first line 
of eq.~(3.39) are unambiguous since all $D$'s \newline ${~~~~~}$ effectively 
anticommute there.}  Note also the identity \cite{hst}
$$ D^i_{(\a}D^j_{\b}D^k_{\g)}=0~.\eqno(3.40)$$

\subsection{6d sBI action in (1,0) superspace}

A (1,0) supersymmetric Maxwell action in 6d superspace is known \cite{nils,hst},
and it reads in our notation as
$$ -\fracmm{1}{4!}\int d^6x \,D^{ij}_{\a\b}W^{\a\b}_{ij} = \int d^6x \,\left\{
-\frac{1}{4}F^2 -\frac{i}{4}\ve^{ij}\j^{\a}_i\pa_{\a\b}\j^{\b}_j-
\frac{1}{4}D_{ij}D^{ij} \right\}~,\eqno(3.41)$$
where $-\frac{1}{4}\int d^6x\, F^2$ is the 6d Maxwell action. The superfield 
action on the left-hand-side of eq.~(3.41) is supersymmetric because of the 
superfield constraint \cite{hst}
$$ D_{\a}^{(i}W^{jk)}_{\b\g}=0 \eqno(3.42)$$
that follows from the defining superspace constraints (3.29) on the off-shell 
(1,0) Maxwell supermultiplet \cite{hst}. Because of eq.~(3.42) the
superfield $W^{ij}_{\a\b}$ is independent upon some of the anticommuting
superspace coordinates (they are linear combinations of $\q$'s). Hence, after
integrating over the rest of 6d superspace coordinates, as in eq.~(3.41), one
arrives at a supersymmetric invariant. The Lorentz and $SU(2)$ invariances
are manifest in eq.~(3.41). This construction is similar to the 
{\it projective} N=2 superspace in 4d \cite{karl,urmala}.  

The superfield  $W^{ij}_{\a\b}$ was identified in ref.~\cite{hst} with the 
(1,0) Maxwell `spin-2' supercurrent superfield. Amongst its $40+40$ components 
are the energy-momentum tensor, the (1,0) supersymmetry current and the 
triplet of conserved $SU(2)$ currents. 

The same reasoning based on eq.~(3.42) further implies that 
$$ \int d^6x\, D^{ijkl}W_{ijkl} ~=~4!\int d^6x\,\det\hat{F} + \ldots,
\eqno(3.43)$$
where the dots stand for the fermionic and $D$-dependent component terms, 
is also a superinvariant that is quartic in the Maxwell field strength. The 
$4\times 4$ determinant in eq.~(3.43) can be expanded in terms of the Lorentz 
invariants (3.15) as
$$ \det\hat{F}= -\frac{1}{4} \left[ 
\tr\hat{F}^4 -\frac{1}{2}(\tr\hat{F}^2)^2\right]~.\eqno(3.44)$$
Eq.~(3.44) is the natural 6d generalization of the 4d Euler-Heisenberg (EH)
lagrangian (2.14), while eq.~(3.43) represents its unique 6d supersymmetric 
generalization. Since the EH lagrangian is just the Maxwell energy-momentum 
tensor squared, its supersymmetric generalization can be naturally understood
as the Maxwell supercurrent superfield squared, both in 4d and 6d.

Eq.~(3.44) is the same quartic combination that appears in the expansion of
the 6d BI action in eq.~(3.19), up to an overall normalization. Hence, 
eqs.~(3.41) and (3.43) with proper normalization are just the leading term
and the `next-to-leading-order-correction', respectively, in the 6d sBI action
that we are looking for. The next (of the 6-th order in $F$) correction to the 
BI action in eq.~(3.19) also has a unique (1,0) supersymmetric extension that
follows the same pattern, namely,
$$ -\fracmm{1}{3\cdot 2^8} \int d^6x\, (D^6)^{\a\b}_{ij}(W^6)^{ij}_{\a\b}~.
\eqno(3.45)$$
This correction is also supersymmetric due to the constraint (3.42). It may
be not accidental that the number (3) of special (non-universal) 
superinvariants given by eqs.~(3.41), (3.43) and (3.45) coincides with the
rank (3) of the 6d Lorentz group. This is related to the fact that the next
superinvariant generalizing the 8-th order terms (in $F$) in the BI action
(3.19) is universal, being given by a full superspace integral. We find
the remarkably simple answer given by
$$ -\fracmm{5}{2^7} \int d^6x\, (D^8)(W^8)=
 -\fracmm{5}{2^7} \int d^6x d^8\q\,W^8~.\eqno(3.46)$$

We thus have a (1,0) supersymmetric generalization of the 6d BI action
$$ S_{\rm bosonic}= -\int d^6x\,\sqrt{-\det(\h_{mn}+F_{mn})} \eqno(3.47)$$
in the form
$$\eqalign{
S[W]~=~& \int d^6x\left\{ - \fracmm{1}{4!}D^{ij}_{\a\b} W^{\a\b}_{ij}
  -\fracmm{1}{2^3\cdot 4!} D^{ijkl}W_{ijkl} \right. \cr
~& \left.  -\fracmm{1}{3\cdot 2^8}(D^6)^{\a\b}_{ij}(W^6)^{ij}_{\a\b} 
-\fracmm{5}{2^7}(D^8)(W^8) + \ldots \right\}~,\cr}\eqno(3.48)$$
where the dots stand for the higher order terms whose component equivalents
are of the 10-th order or higher in $F$. 

From the group-theoretical viewpoint, the Lorentz invariance of eq.~(3.48) is
manifest, being related to the existence of a trivial representation in the
$SU(4)$ product 
$$\un{6}\times \un{6} =\un{1} + \un{15} + \un{20'}~,\eqno(3.49)$$
and similarly for the $SU(2)$ irreps,
$$ \eqalign{
\un{3}\times \un{3} =& \un{1} +\un{3} +\un{5}~,\cr
\un{4}\times \un{4} =& \un{1} +\un{3} +\un{5} +\un{7}~.\cr}
\eqno(3.50)$$ 

Being the full superspace integral, the `top' superinvariant of eq.~(3.46) is
already manifestly supersymmetric {\it without} the use of the constraint 
(3.42). Hence, similarly to the previously considered cases (sect.~2), it can
be further generalized by inserting a structure function $\cz(K,L,M)$ at our
disposal into the superfield lagrangian,
$$ 
\int d^6x d^8\q\,W^8~\to~ \int d^6x d^8\q\,\cz(D^2W^2,D^4W^4,D^6W^6)W^8~,
\eqno(3.51)$$
where we have introduced the new bosonic scalar superfields $K,L$ and $M$ as
the supercovariant derivatives of $W$:
$$\eqalign{
D^2W^2 ~&\sim~  - \fracmm{1}{4!}D^{ij}_{\a\b} W^{\a\b}_{ij}\equiv K~,\cr
D^4W^4 ~&\sim~ -\fracmm{1}{2^3\cdot 4!} D^{ijkl}W_{ijkl}\equiv L~,\cr
D^6W^6 ~&\sim~   -\fracmm{1}{3\cdot 2^8}(D^6)^{\a\b}_{ij}(W^6)^{ij}_{\a\b}
\equiv M~.\cr}\eqno(3.52)$$
The $W^8$ factor in eq.~(3.51) `soaks up' the anticommuting spinor 
derivatives in the full superspace measure, so that the structure function 
$\cz(K,L,M)$ subject to the `initial' condition $\cz(0,0,0)=1$ can only affect
the terms of the 
order $F^{10}$ or higher in the component expansion of the action (3.51). 

The $F$-products (3.15) are simply related to those defined by eq.~(3.52),
namely,
$$ \eqalign{
\fracmm{1}{2^3}\tr\hat{F}^2 ~=~ & -\left.K\right|~,\cr
\fracmm{1}{2^5}\tr\hat{F}^4 ~=~ & \left.(L+K^2)\right|~,\cr
\fracmm{1}{3\cdot 2^7}\tr\hat{F}^6 ~=~ & 
-\left. (M+\frac{3}{2}KL+\frac{1}{3}K^3)\right|~,\cr}
\eqno(3.53)$$
where $\left.\right|$ means taking merely the $F$-dependent terms in the first
component of a superfield. It is now straightforward ro rewrite eqs.~(3.18) and
(3.19) to another form, in terms of the new variables $K$, $L$ and $M$. We find
$$ -\det(\h_{mn}+F_{mn})=1+ 2(K+L+M)+2KL +K^2~,\eqno(3.54)$$
and
$$ \sqrt{ -\det(\h_{mn}+F_{mn})}=1+(K+L+M)-\frac{1}{2}(L^2+2KM)\cz~.
\eqno(3.55)$$
Hence, the structure function $\cz(K,L,M)$ reads
$$ \cz=\fracmm{ 1+(K+L+M) -\sqrt{1+ 2(K+L+M)+2KL +K^2}}{KM+\frac{1}{2}L^2}
=1-K+\ldots~, \eqno(3.56)$$
where the dots stand for the higher order terms (quartic in $F$ or higher) in 
the expansion of the exact (non-perturbative) formula on the left-hand-side.

We are now in a position to write down the full (1,0) supersymmetric 
Born-Infeld-Goldstone action describing the gauge-fixed 6d `spacetime-filling'
D-5-brane in 6d superspace as 
$$\eqalign{
S[W]~=~& \int d^6x\,\left\{ - \fracmm{1}{4!}D^{ij}_{\a\b} W^{\a\b}_{ij}
  -\fracmm{1}{2^3\cdot 4!} D^{ijkl}W_{ijkl}  -\fracmm{1}{3\cdot 2^8}
(D^6)^{\a\b}_{ij}(W^6)^{ij}_{\a\b} \right\} \cr
~ & - \fracmm{5}{3\cdot 2^7}\int d^6x d^8\q\,\cz(D^2W^2,D^4W^4,D^6W^6)W^8~,
\cr}\eqno(3.57)$$
where the structure function $\cz$ is given by eq.~(3.56). The 6d action (3.57)
has linearly realised (1,0) supersymmetry, whereas the non-linear structure of
the function $\cz$ in eq.~(3.56) is supposed to be dictated by yet another 
non-linearly realised (1,0) supersymmetry which is hidden, being spontaneously 
broken 
({\it cf.} sect.~2). The $(2,0)$ supersymmetry algebra in 6d reads \cite{hst}
$$  \{ Q^I_{\a}, Q^J_{\b}\} = i\O^{IJ}\pa_{\a\b}~,\eqno(3.58)$$
where $\O^{IJ}$ is the invariant metric of $USp(4)$ and $I,J=1,2,3,4$. It is
straightforward to calculate the transformation laws of the hidden, 
spontaneously broken (1,0) supersymmetry by using the general theory of 
non-linear realisations \cite{pro}. We would like to emphasize here that the
corresponding Goldstone-Maxwell action (3.57) was found by a direct 
supersymmetrization of the 6d bosonic Born-Infeld action that already `knows' 
about partial supersymmetry breaking due to its BPS nature. The action (3.57) 
also gives the 6d superspace action of the gauge-fixed D-5-brane whose 
worldvolume is given by the whole 6d spacetime (sect.~1).

A plain dimensional reduction of the 6d super-Born-Infeld action (3.57) down to
4d leads to the 4d, N=2 supersymmetric Born-Infeld-Nambu-Goto-type action 
considered in subsect.~2.3,
$$ W^{\a}_i~\to~D^{\a}_iW~,\qquad \a=1,2~.\eqno(3.59)$$

Since the 6d superspace action (3.57) is written down entirely in terms of 
the 6d superfield strength $W^{\a}_i$, the same action in components is going
to be dependent upon the 6d Maxwell vector field $A_m$ only via its field 
strength $F_{mn}=\pa_mA_n-\pa_nA_m$. Hence, after the dimensional reduction,
$$ \pa_4=\pa_5=0~,\quad{\rm and}\quad A_4+iA_5=P+iQ~,\eqno(3.60)$$
the resulting 4d, N=2 super-BING action will be dependent upon the 4d scalars
$(P,Q)$ only via their 4d spacetime derivatives, $\pa_{\m}P$ and  $\pa_{\m}Q$.
It guarantees the rigid off-shell symmetry    
$$ P\to P +const.,\quad{\rm and}\quad  Q\to Q +const.,\eqno(3.61)$$
which is necessary for the Goldstone interpretation of the real scalars $(P,Q)$
as the Goldstone bosons associated with spontaneously broken translations. The
symmetry (3.61) is not manifest in our 4d, N=2 super-BING action (2.42), but 
it becomes manifest after rewriting it to the more symmetric 6d form as above
({\it cf.} ref.~\cite{rte}). 

When using the identities
$$ \eqalign{
W_{ijkl}~=~& 2W_{ij}^{\a\b}W_{kl\a\b}~,\cr
(W^6)^{ij}_{\a\b}~=~&\frac{1}{6}W_{\a\b kl}W^{ijkl}~,\cr
(W^8)~=~&\frac{1}{48}W_{ij}^{\a\b}(W^6)^{ij}_{\a\b}~,\cr}
\eqno(3.62)$$
and similarly for the products (3.39) of the 6d superspace supercovariant 
derivatives $D^{ij}_{\a\b}$ (modulo total derivatives in 6d spacetime), it is 
possible to rewrite the action (3.57) to the form of a 
`non-linear sigma-model', {\it viz.}
$$\eqalign{
S[W]~=~& -\fracmm{1}{4!}\int d^6x\, D^{ij}_{\a\b}\left\{ 
W^{\a\b}_{ij} +\fracmm{1}{2}D^{\a\b}_{kl}\left(W_{ij}^{\g\d}W_{\g\d}^{kl}
\right)\right. \cr
& \left.+
\fracmm{1}{2^6\cdot 3^2}D_{ijkl}\left(W_{mn}^{\a\b}W^{mn\g\d}W_{\g\d}^{kl}
\right) + \fracmm{5}{2^{12}\cdot 3^4}(D^6)^{\a\b}_{ij}\left[ 
W^{kl}_{\g\d}W^{\s\t}_{kl}W^{mn}_{\s\t}W^{\g\d}_{mn}\cz\right]\right\} \cr
~\equiv~&  -\fracmm{1}{4!}\int d^6x\, D^{ij}_{\a\b}\,X^{\a\b}_{ij}~,\cr}
\eqno(3.63)$$
where we have introduced the {\it improved} Maxwell-Goldstone (1,0)
supercurrent superfield $X^{\a\b}_{ij}\equiv \left(W^{\a\b}_{ij}\right)_{\rm
improved}\,$. The latter seems to satisfy to the very simple non-linear 
superfield constraint~\footnote{We checked it in a few leading 
orders in $X$.}
$$ X^{\a\b}_{ij}=\fracm{1}{2}D^{\a\b}_{kl}
\left(X^{\g\d}_{ij}X_{\g\d}^{kl} \right) +W^{\a\b}_{ij}~.\eqno(3.64)$$
This off-shell superfield constraint is the 6d superspace generalization of 
the similar 4d superspace constraints in eqs.~(2.29) and (2.44).
\vglue.2in

\section{Conclusion}

In this paper we advocate the extended superspace approach for constructing
some gauge-fixed supersymmetric D-brane actions that are given by 
superextensions of the Born-Infeld (or Born-Infeld-Nambu-Goto) actions. Those
actions are the Goldstone actions associated with partial spontaneous breaking
of extended supersymmetry with 16 supercharges down to 8 supercharges in four
and six spacetime dimensions. We believe that the extended superspace 
approach is adequate for these purposes, since it is (i) simple, 
(ii) transparent, (iii) universal and (iv) powerful. The number (8) of
unbroken supercharges is the maximal one allowed in the conventional 
{\it off-shell} superspace formulation of supersymmetric field theories.  

The main results of our investigation are given by the 6d, (1,0) 
supersymmetric Born-Infeld-Goldstone action (subsect.~3.3) and its 4d, N=2
supersupersymmetric counterpart (subsect.~2.3),~\footnote{The 4d, N=2 
supersymmetric BING action (2.42) was found for the first time in 
ref.~\cite{k2}.} as well as their `non-linear sigma-model' representations 
given by eqs.~(3.63) and (3.64), and eqs.~(2.43) and (2.44), respectively. 
In our approach, 
the irreducibility of the Goldstone vector supermultiplet is ensured by the
standard `linear' off-shell superfield constraints on the Maxwell-Goldstone 
superfield, whereas the non-linearity of the Born-Infeld-Goldstone action is 
represented by the off-shell superfield structure functions or, equivalently,
the `non-linear sigma-model' off-shell superfield constraints 
({\it cf.} refs.~\cite{bik,rte}). Despite of the presence of higher 
derivatives to all orders, all the actions considered have no ghosts and lead
to causal propagation of the physical fields. Moreover, they enjoy the
auxiliary freedom, i.e. their auxiliary fields do not propagate, being 
vanishing on-shell. The 4d, N=2 super-BING action is self-dual with respect 
to the N=2 supersymmetric electric-magnetic duality. 

Other Goldstone actions associated with different  patterns of partial
supersymmetry breaking $N=2\to N=1$ in 4d are known to be related to other
massless Goldstone (chiral or tensor) N=1 supermultiplets \cite{bag,stony,rte}.
The Goldstone action associated with partial supersymmetry breaking 
$(1,1)\to(1,0)$ in 6d, with a tensor (1,0) supermultiplet of Goldstone 
fields, is known to be the (gauge-fixed) effective field theory action in the 
M-5-brane worldvolume \cite{blnpst,apps}. Amongst the components of
the (1,0) tensor multiplet in 6d, there is a gauge two-form whose field 
strength is self-dual. After dimensional reduction on a torus, this self-dual 
field yields a 4d Maxwell gauge field, while the dimensionally reduced 
M-5-brane effective action appears to be the Born-Infeld-Goldstone action in 
4d \cite{schpe}. This way of doing also allows one to make manifest the 
electric-magnetic self-duality of the gauge-fixed D-3-brane action, and extend 
it further to a full classical $SL(2,{\bf R})$ duality, with the background 
axion-dilaton fields being taken into account \cite{tse2}. The $SL(2,{\bf Z)}$
self-duality expected to survive in quantum theory then appears to be related to the 
compactification (torus) geometry, being identified with the invariance of the torus 
under the $SL(2,{\bf Z})$ transformations of its complex structure \cite{verlinde,berman}. 
It also implies that the supersymmetric version of the dimensionally reduced (and truncated) 
theory to be obtained from the 6d self-interacting (1,0) tensor multiplet action, when 
only two Goldstone bosons are kept, should be given by our 4d, N=2 
supersymmetric BING (or Goldstone-Maxwell) action (subsect.~2.3).

The self-duality condition attached to the on-shell 6d tensor (1,0) multiplet 
makes it difficult to find its 6d Lorentz-invariant and supersymmetric action 
that could be useful for doing quantum calculations. Our 6d Goldstone-Maxwell 
superfield action is manifestly 6d Lorentz invariant and (1,0) supersymmetric,
while it can be quantized directly in 6d superspace without obstructions. As a
by-product of our 6d superfield analysis, we found the full list of 
non-universal (1,0) superinvariants on the (1,0) Maxwell superfield 
constraints in 6d superspace. Those higher-derivative supersymetric and gauge 
invariants are likely to be forbidden as possible UV counterterms, but they may
still appear as finite local corrections to the low-energy effective actions of quantum 
supersymmetric interacting gauge field theories in six dimensions.

\newpage

\section*{Acknowledgements}

I am grateful to B. de Wit, B. E. W. Nilsson, E. Bergshoeff and H. Nicolai
for their kind hospitality extended to me at the Institute for Theoretical
Physics in Utrecht, the Institute for Theoretical Physics in G\"oteborg,
the Institute for Theoretical Physics in Groningen and the 
Albert-Einstein-Institute in Potsdam, respectively, where this investigation
was done. I would like to thank J. Bagger for informing me about  
ref.~\cite{bg}, J. Pleba\'nski for sending me a copy of ref.~\cite{pleb},
and D. Berman and A. Westerberg for correspondence. I also acknowledge 
discussions with S. J. Gates Jr., E. A. Ivanov and A. A. Tseytlin at the early 
stage of this work.


\newpage

\begin{thebibliography}{99}

\bibitem{aps} M. Aganagic, C. Popescu and J. H. Schwarz, 
\np{495}{97}{99}
\bibitem{nil} M. Cederwall, A. von Gussich, B. E. W. Nilsson and 
A. Westerberg, \\ \np{490}{97}{163};\\
M. Cederwall, A. von Gussich, B. E. W. Nilsson, P. Sundell and 
A. Westerberg, \\ \np{490}{97}{179}
\bibitem{ber} E. Bergshoeff and P. K. Townsend, \np{490}{97}{145}
\bibitem{sor} I. A. Bandos, D. P. Sorokin and M. Tonin, 
\np{497}{97}{275}
\bibitem{kal} R. Kallosh, {\it Volkov-Akulov theory and D-branes}, Talk
given at Intern. Seminar on Supersymmetry and Quantum Field Theory,
Kharkov, Ukraine, 5--7 Jan., 1997, in `Kharkov 1997. Supersymmetry 
and Quantum Field Theory', p.~49--58; hep-th/9705118
\bibitem{polbook} J. Polchinski, `String Theory', in two volumes,
Cambridge University Press, 1998 
\bibitem{town} P. Townsend, {\it Four lectures about M theory}, 
Talks given at Summer School in High-Energy Physics and Cosmology, 
Trieste, Italy, 10 June - 26 July 1996, in `Trieste 1996. High Energy 
Physics and Cosmology', p.~385-438; hep-th/9612121 
\bibitem{av} D. V. Volkov and V. P. Akulov, \pl{46}{73}{109}
\bibitem{bagger} J. Bagger and A. Galperin, {\it Linear and non-linear
supersymmetries}, Talk given at Intern. Seminar on Supersymmetries and 
Quantum Symmetries, Dubna, Russia, 22--26 July 1997; hep-th/9810109
\bibitem{bik} S. Bellucci, E. Ivanov and S. Krivonos, {\it Partial breaking
N=4 to N=2: hypermultipet as a Goldstone superfield}, Talks given at 11th 
Intern. Conference on Problems of Quantum Field Theory, Dubna, Russia,
13--17 July 1998, and 32nd Intern. Symposium Ahrenshoop on the Theory of 
Elementary Particles, Buckow, Germany, 1-5 Sept., 1998, hep-th/9809190; and 
{\it Partial breaking of N=1, D=10 supersymmetry}, hep-th/9811244 
\bibitem{kle} I. R. Klebanov, \np{496}{97}{231};\\
S. S. Gubser, I. R. Klebanov and A. A. Tseytlin, \np{499}{97}{217}
\bibitem{agat} S. J. Gates, Jr., \pl{365}{96}{132}, \np{485}{97}{145}
\bibitem{swer} A. Karlhede, U. Lindstr\"om, M. Ro\v{c}ek and
G. Theodoridis, \np{294}{87}{498}
\bibitem{bi} M. Born and L. Infeld, Proc. Roy. Soc. (London) {\bf A144} 
(1934) 425;\\
M. Born, Ann. Inst. Poincar\'e {\bf 7} (1939) 155
\bibitem{schr} E. Schr\"odinger, Proc. Roy. Soc. (London) {\bf A150} 
(1935) 465
\bibitem{gzu} M. K. Gaillard and B. Zumino, {\it Self-duality in non-linear
electromagnetism}, Talk given at Intern. Seminar on Supersymmetry and Quantum 
Field Theory, Kharkov, Ukraine, 5--7 Jan., 1997, in `Kharkov 1997. 
Supersymmetry and Quantum Field Theory', p.~121--129, hep-th/9705226; 
{\it Non-linear electromagnetic self-duality and Legendre transformations}, 
Contributed to a Newton Institute Euroconference on Duality and 
Supersymmetric Theories, Cambridge, England, 7--18 April 1997, hep-th/9712103
\bibitem{pleb} J. Pleba\'nski, Nordita Lectures on Non-Linear Electrodynamics,
Copenhagen, October 1968
\bibitem{gr} G. W. Gibbons and D. A. Rasheed, \np{454}{95}{185}
\bibitem{hlp} J. Hughes and J. Polchinski, \np{278}{86}{147};\\
J. Hughes, J. Liu and J. Polchinski, \pl{180}{86}{370}
\bibitem{strings} E. Bergshoeff, E. Sezgin, C. N. Pope and P. K. Townsend, 
\pl{188}{87}{70};\\
R. R. Metsaev, M. A. Rachmanov and A. A. Tseytlin, \pl{193}{87}{207};\\
A. Abouelsaood, C. G. Callan, C. R. Nappi and S. A. Yost, 
\np{280}{87}{599};\\
R. G. Leigh, \mpl{4}{89}{2073}
\bibitem{eh} W. Heisenberg and H. Euler, Z. Phys. {\bf 98} (1936) 714 
\bibitem{bg} J. Bagger and A. Galperin,  \pr{55}{97}{1091}
\bibitem{rte} M. Ro\v{c}ek and A. A. Tseytlin, {\it Partial breaking of
global D=4 supersymmetry, constrained superfields, and 3-brane actions},
hep-th/9811232
\bibitem{stony} F. Gonzalez-Rey, I. Y. Park and M. Ro\v{c}ek, {\it On dual
3-brane actions with partially broken N=2 supersymmetry}, hep-th/9811130 
\bibitem{cwz} S. Coleman, J. Wess and B. Zumino, Phys. Rev. {\bf 177} (1969)
2239;\\
C. G. Callan Jr., S. Coleman, J. Wess and B. Zumino, Phys. Rev. {\bf 177} 
(1969) 2247;\\
D. V. Volkov, Sov. J. Particles and Nuclei {\bf 4} (1973) 3;\\
V. I. Ogievetsky, in Proceedings of X-th Winter School of Theoretical Physics
in Karpacz, Poland; Wrozlaw (1974), Vol.~1, p.~227 
\bibitem{k2} S. V. Ketov, {\it A manifestly N=2 supersymmetric Born-Infeld
action}, hep-th/9809121
\bibitem{tse2} A. A. Tseytlin, \np{469}{96}{51}
\bibitem{hz} M. Abou Zeid and C. M. Hull, \pl{404}{97}{264}, \ibid{428}{98}{277} 
\bibitem{cf} S. Cecotti and S. Ferrara, \pl{187}{86}{335}
\bibitem{bw} J. Wess and J. Bagger, {\it Supersymmetry and Supergravity},
Princeton University Press, 1992
\bibitem{di} M. J. Duff and C. J. Isham, \pl{86}{79}{157}
\bibitem{apt} I. Antoniadis, H. Partouche and T. R. Taylor, \pl{372}{96}{83}
\bibitem{wess} J. Wess, Acta Physica Austr. {\bf 41} (1975) 409
\bibitem{oldrev} S. V. Ketov, Fortschr. Phys. {\bf 36} (1988) 361
\bibitem{sg} S. J. Gates Jr., and W. Siegel, \np{189}{81}{295}
\bibitem{buch} I. L. Buchbinder, E. I. Buchbinder, E. A. Ivanov, S. M.
Kuzenko and B. A. Ovrut, \pl{412}{97}{309}
\bibitem{krev} S. V. Ketov, {\it Analytic tools to brane technology in N=2
gauge theories mith matter}, hep-th/9806009; to appear in Fortschr. Phys.
\bibitem{mez} L. Mezincescu, {\it On the superfield formulation of O(2)
supersymmetry}, Dubna preprint P2--12572, 1979 (unpublished)
\bibitem{iz} E. A. Ivanov and B. M. Zupnik, {\it Modified N=2 supersymmetry 
and Fayet-Iliopoulos terms}, hep-th/9710236
\bibitem{pro} work in progress
\bibitem{gw} D. Gross and E. Witten, \np{277}{86}{1}
\bibitem{gv} S. J. Gates, Jr. and Sh. Vashakidze, \np{291}{87}{172} 
\bibitem{brs} E. Bergshoeff, M. Rakowski and E. Sezgin, 
\pl{185}{87}{371}
\bibitem{hst} P. S. Howe, G. Sierra and P. K. Townsend, \np{221}{83}{331}
\bibitem{van} P van Nieuwenhuizen, Phys. Rep. {\bf 68} (1981) 189 
\bibitem{aw} L. Alvarez-Gaum\`e and E. Witten, \np{234}{84}{269}
\bibitem{kano} S. V. Ketov, \cqg{7}{90}{1387}
\bibitem{nils} B. E. W. Nilsson, \np{174}{80}{335}
\bibitem{kto} T. Kugo and P. K. Townsend, \np{221}{83}{357}
\bibitem{kty} S. V. Ketov and I. V. Tyutin, JETP Lett. {\bf 39} (1984) 
703, and Theor.~Math.~Phys. {\bf 61} (1984) 1126
\bibitem{karl} A. Karlhede, U. Lindstr\"om  and M. Ro\v{c}ek, 
\pl{147}{84}{297}
\bibitem{urmala} S. V. Ketov, {\it Self-interaction for N=2 hypermultiplets in
d=4, and ultra-violet finiteness of N=2 non-linear sigma-models in d=2}, in
the Proceedings of the Intern. Seminar on Group Theory Methods in Physics, 
Urmala, USSR, 1985; Moscow: Nauka, Vol.~1, p.~87
\bibitem{bag} J. Bagger and A. Galperin, \pl{336}{94}{25}, \ibid{412}{97}{296}
\bibitem{blnpst} I. Bandos, K. Lechner, A. Nurmagambetov, P. Pasti, D. Sorokin
and M. Tonin, \prl{78}{97}{4332}, \pl{408}{97}{135} 
\bibitem{apps} M. Aganagic, J. Park, C. Popescu and J. H. Schwarz, 
\np{496}{97}{191}
\bibitem{schpe} M. Perry and J. H. Schwarz, \np{489}{97}{47}
\bibitem{verlinde} E. Verlinde, \np{455}{95}{211}
\bibitem{berman} D. Berman, \pl{403}{97}{250}, \ibid{409}{97}{153}. 

\end{thebibliography}
\end{document}